\documentclass[a4paper,twoside]{article}
\newcommand{\affil}[1]{$^{\rm #1}$}
\usepackage[authoryear]{natbib}
\bibpunct{(}{)}{;}{a}{}{,}
\usepackage{graphicx}
\usepackage{amsmath} %allows 'aligns' rather than equations
\usepackage{lscape} %landscape tables
\date{} %Please leave the date blank
\newcommand{\etal}{\hbox{et al.}} 
\newcommand{\vlsr}{\hbox{v$_{\textrm{\tiny{LSR}}}$}}
\newcommand{\Trms}{\hbox{T$_{\textrm{\tiny{RMS}}}$}}
\newcommand{\fwhm}{\hbox{$\Delta$v$_{\textrm{\tiny{FWHM}}}$}}
\newcommand{\Tpeak}{\hbox{T$_{\textrm{\tiny{Peak}}}$}}

\newcommand{\Nh}{\hbox{$\overline{N_{H_2}}$}}

\newcommand{\RXJ}{RX\,J1713.7$-$3946}

\title{\large\bf\flushleft Dense Gas Towards the \RXJ\ Supernova Remnant}
\author{\parbox{\textwidth}{\flushleft
\vspace{-0.5cm}
{\it Nigel I. Maxted\affil{A}, Gavin P. Rowell\affil{A}, Bruce R. Dawson\affil{A},  Michael G. Burton\affil{B}, Yasuo Fukui\affil{C}, Jasmina Lazendic\affil{D}, Akiko Kawamura\affil{C}, Hirotaka Horachi\affil{C}, Hidetoshi Sano\affil{C}, Andrew J. Walsh\affil{E}, Satoshi Yoshiike\affil{C} and Tatsuya Fukuda\affil{C}\\}
\vspace{0.4cm}
{\small \affil{A}\,School of Chemistry \& Physics, University of Adelaide, Adelaide, 5005,  Australia}\\
{\small \affil{B}\,School of Physics, University of New South Wales, Sydney, 2052, Australia}\\
{\small \affil{C}\,Department of Astrophysics, Nagoya University, Furocho, Chikusa-ku, Nagoya, Aichi, 464-8602, Japan}\\
{\small \affil{D}\,School of Physics, Monash University, Melbourne, 3800, Australia}\\
{\small \affil{E}\,International Centre for Radio Astronomy Research, Curtin University, GPO Box U1987, Perth, Australia}\\
}}
\begin{document}
\twocolumn[
\begin{changemargin}{.8cm}{.5cm}
\begin{minipage}{.9\textwidth}
\vspace{-1cm}
\maketitle
%
%
%%%%%%%%%%%%%     ABSTRACT    %%%%%%%%%%%%%
%Abstract of no more than 200 words here.
\small{\bf Abstract:} We present results from a Mopra 7\,mm-wavelength survey that targeted the dense gas-tracing CS(1-0) transition towards the young $\gamma$-ray-bright supernova remnant, \RXJ\ (SNR G\,347.3$-$0.5). % The Mopra radio telescope was used to target the high critical density tracer CS(1$-$0), complementing previous Nanten2 molecular gas studies of CO transitions.
In a hadronic $\gamma$-ray emission scenario, where cosmic ray protons interact with gas to produce the observed $\gamma$-ray emission, the mass of potential cosmic ray target material is an important factor. We summarise newly-discovered dense gas components, towards Cores G and L, and Clumps N1, N2, N3 and T1, which have masses of 1\,-\,10$^4$\,M$_{\odot}$. We argue that these components are not likely to contribute significantly to $\gamma$-ray emission in a hadronic $\gamma$-ray emission scenario. This would be the case if \RXJ\ were at either the currently favoured distance of $\sim$1\,kpc or an alternate distance (as suggested in some previous studies) of $\sim$6\,kpc.

%The mass of potential cosmic ray target material may be an important factor in hadronic $\gamma$-ray emission scenarios (p-p interactions), but newly-discovered dense gas components, may not significantly contribute to this component if it exists. This would be the case even if \RXJ\ were not at the currently favoured distance of $\sim$1\,kpc, but instead $\sim$6\,kpc. %Background gas in the Norma and 3\,kpc-expanding arms display dense gas components.

This survey also targeted the shock-tracing SiO molecule. Although no SiO emission corresponding to the \RXJ\ shock was observed, vibrationally-excited SiO(1-0) maser emission was discovered towards what may be an evolved star. Observations taken one year apart confirmed a transient nature, since the intensity, line-width and central velocity of SiO(J=1-0,v=1,2) emission varied significantly.

%The shock-tracing SiO molecule did not provide millimetre evidence for the existence of a supernova shock associated with molecular gas, but vibrationally-excited SiO(1-0) maser emission was discovered towards what may be an evolved star. Observations taken one year apart confirm a transient nature, with the intensity, line-width and central velocity of SiO(J=1-0,v=1,2) emission significantly varying.

%%%%%%%%%%%%%     KEYWORDS    %%%%%%%%%%%%%
\medskip{\bf Keywords:} molecular data - supernovae: individual: RX\,J1713.7$-$3946 - ISM: clouds - cosmic
rays - gamma rays: ISM
% Please write all keywords in lower case. PASA uses the
% standard list of subject headings adopted by The Astrophysical Journal
% and available from http://www.journals.uchicago.edu/ApJ/keywords_text.html.
% Keywords are separated by em-dashes, i.e. ---

%%%%%%%%DO NOT EDIT%%%%%%%%%%%%
\medskip
\medskip
\end{minipage}
\end{changemargin}
]
\small
%%%%%%%%EDIT FROM HERE%%%%%%%%%%%%

\section{Introduction}\label{sec:intro}
\RXJ\ (G\,347.3$-$0.5) is a supernova remnant (SNR) that is bright in X-ray emission \citep{Pfeffermann:1996,Cassam:2004,Acero:2009} and is one of the brightest sources in the TeV $\gamma$-ray sky \citep{Aharonian:2006,Aharonian:2007}. This remnant is therefore ideal for investigating the possibility of acceleration of cosmic rays (CRS) in the shocks of SNRs.
\begin{figure}[!h]
\centering
\includegraphics[width=0.41\textwidth, angle =90]{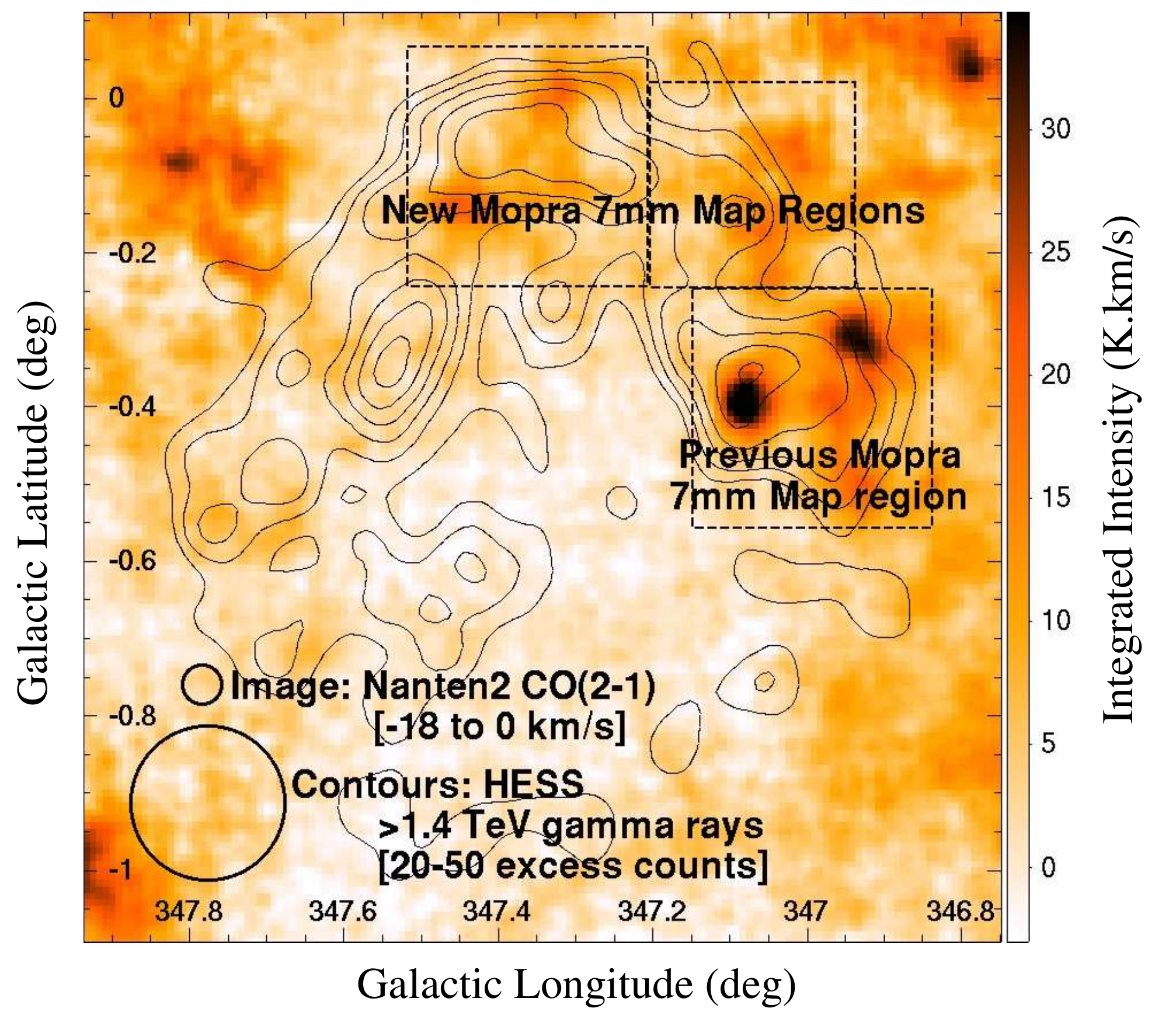}
\caption{Integrated Nanten2 CO(2-1) emission image \citep{Fukui:2008} of gas believed to be associated with \RXJ\ (\vlsr$=-$18~to~0\,km\,s$^{-1}$) with overlaid HESS $>$1.4\,TeV $\gamma$-ray photon excess count contours \citep{Aharonian:2007}. Three square regions (one, the southern-most, from a previous investigation) indicate the extent of Mopra 7\,mm mapping carried out in this study. \label{fig:CO_wHESS}}
\end{figure}
\begin{figure}[!h]
\centering
\includegraphics[width=0.40\textwidth, angle =90]{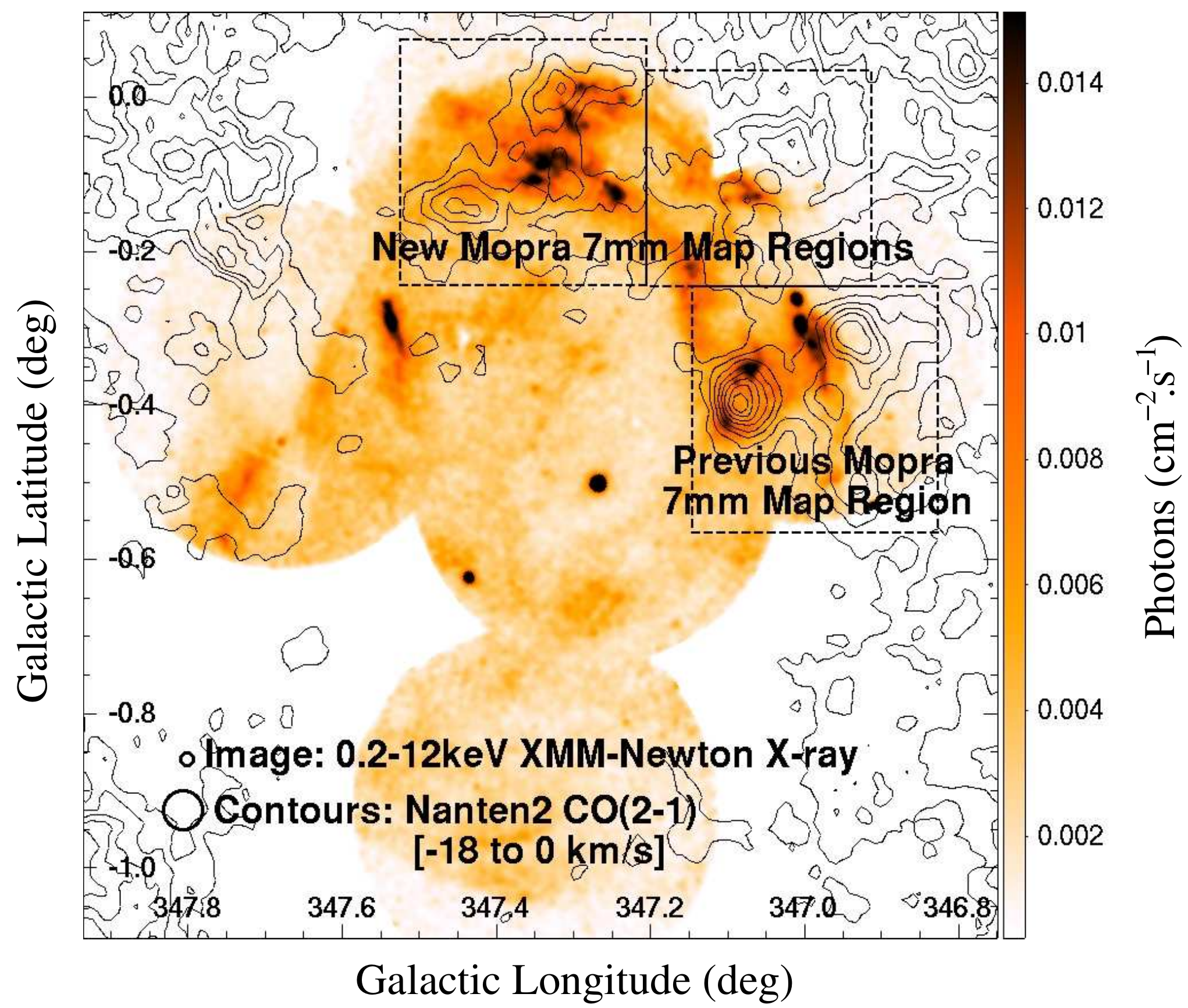}
\caption{An XMM-Newton 0.5-4.5\,keV X-ray image \citep{Acero:2009} with overlaid integrated Nanten2 CO(J=2-1) emission (\vlsr$=-$18~to~0\,km\,s$^{-1}$) contours. CO(J=2-1) contours span 5\,K\,km\,s$^{-1}$ to 40\,K\,km\,s$^{-1}$ in increments of 5\,K\,km\,s$^{-1}$. The XMM-Newton image has been exposure-corrected and smoothed with a Gaussian of FWHM=30$^{\prime \prime}$. Three square regions (one, the southern-most, from a previous investigation) indicate the extent of Mopra 7\,mm mapping carried out in this study. \label{fig:CO_xmm}}
\end{figure}
%However, a hadronic component may be possible towards dense cloud clumps \citep{Zirakashvili:2010}, 
% that may favour lepton-dominated $\gamma$-ray models (ie. inverse Compton scattering of TeV electrons) \citep{Porter:2006,Aharonian:2007,Berezhko:2010,Ellison:2010,Zirakashvili:2010}. However, a hadronic component may be possible towards dense cloud clumps \citep{Zirakashvili:2010}, and the

Knowledge of the distribution of matter towards RX\,J1713.7$-$3946 is important to help distinguish between models of $\gamma$-ray emission which are dominated by high energy electrons (inverse-Compton scattering of photons) and models dominated by CR hadrons (p-p interactions). The latter scenario requires target material for CRs. CR target material may be in any chemical form, including molecular and atomic gas, which have been studied with CO \citep{Fukui:2003,Moriguchi:2005,Fukui:2008} and HI (+CO) \citep{Fukui:2012}, respectively.

We take particular interest in the study of dense molecular gas, which is well-traced by CS(1-0) emission (critical density $\sim$10$^5$\,cm$^{-3}$). This transition highlights gas mass possibly missed by other tracers (such as CO) and aids in the prediction of possible small-scale $\gamma$-ray emission features caused by energy dependent CR diffusion into dense gas (e.g. \citeauthor{Gabici:2009}, \citeyear{Gabici:2009}; \citeauthor{Casanova:2010}, \citeyear{Casanova:2010}; \citeauthor{Fukui:2012}, \citeyear{Fukui:2012}; \citeauthor{Maxted:2012}, \citeyear{Maxted:2012}).

The latest Fermi-LAT observations \citep{Abdo:2011} towards \RXJ\ exhibit a low but hard-spectrum flux of 1-10\,GeV $\gamma$-ray emission uncharacteristic of previous hadronic models %, being more consistent with lepton-dominated gamma-ray models (ie. inverse Compton scattering of TeV electrons)
\citep{Porter:2006,Aharonian:2007,Berezhko:2010,Ellison:2010,Zirakashvili:2010}. However, if one considers an inhomogeneous ISM into which the SNR shock has expanded, the existence of a hadronic component may still be plausible \citep{Zirakashvili:2010,Inoue:2012}. There is also support for such a scenario when considering the existence of additional atomic gas together with the molecular \citep{Fukui:2012}.

\RXJ\ exhibits a shell-like structure at keV X-ray and TeV $\gamma$-ray energies (see Figures\,\ref{fig:CO_wHESS} and \ref{fig:CO_xmm}), with the keV X-ray emission best corresponding to a void in molecular gas, bordering molecular cores at a kinematic distance of $\sim$1\,kpc (\vlsr$\sim-$10\,km\,s$^{-1}$) \citep{Fukui:2003,Moriguchi:2005}. This distance is consistent with X-ray absorption modeling by \citet{Pfeffermann:1996} for a plausible SNR-age ($\sim$1600\,years old) that corresponds to a progenitor event observed by Chinese astronomers in 393\,AD \citep{Wang:1997}.

\citet{Fessen:2012} argue that if a SN\,393\,AD connection to RX\,J1713.7$-$3946 is assumed, the initial \RXJ\ SN explosion must have been optically subluminous for the $\sim$1\,kpc distance to hold. Irrespective of this, recent work still favours a $\sim$1\,kpc distance (eg. \citeauthor{Inoue:2012}, \citeyear{Inoue:2012}, \citeauthor{Fukui:2012}, \citeyear{Fukui:2012}), and an examination of keV X-ray emission (Figure\,\ref{fig:CO_xmm}) suggests some degree of anticorrelation with CO(2-1) peaks at \vlsr$\sim -$10\,km\,s$^{-1}$ (a kinematic distance of $\sim$1\,kpc). \citet{Sano:2010} note that synchrotron intensity peaks on the boundary of the molecular clump, Core\,C may suggest a compression triggered by the \RXJ\ shock. Similarly, the edge of Core\,D is coincident with an X-ray intensity peak, consistent with shock compression at a distance of $\sim$1\,kpc. Furthermore, the northern and western regions of RX J1713.7$-$3946, where cores D and C are located, correspond to peak fluxes of TeV $\gamma$-ray emission. Good gas-$\gamma$-ray overlap, like that seen towards \RXJ\ and the $\sim$1\,kpc gas, is indeed expected in a hadronic scenario for $\gamma$-ray emission.

%There are, however, some difficulties connecting the 393\,AD event with the currently-favoured distance ($\sim$1\,kpc) of \RXJ . \citet{Fessen:2012} argue that if a SN\,393\,AD connection is assumed, the initial \RXJ\ SN explosion must have been subluminous for the $\sim$1\,kpc distance to hold. But since the \RXJ\ $\gamma$-ray flux above 1\,TeV is $\sim$66\% that of the Crab nebula, if \RXJ\ was at the other distance suggested in literature, $\sim$6\,kpc, one might consider \RXJ\ to be overluminoius in $\gamma$-rays in the present. Irrespective of this, recent work still favours a $\sim$1\,kpc distance (eg. \citeauthor{Inoue:2012}, \citeyear{Inoue:2012}, \citeauthor{Fukui:2012}, \citeyear{Fukui:2012}), and an examination of keV X-ray emission (Figure\,\ref{fig:CO_xmm}) suggests some degree of anticorrelation with CO(2-1) peaks at \vlsr$\sim -$10\,km\,s$^{-1}$. \citet{Sano:2010} note that synchrotron intensity peaks on the boundary of Core\,C may suggest a compression triggered by the \RXJ\ shock. Similarly, the border of Core\,D is coincident an X-ray intensity peak.

In a previous study \citep{Maxted:2012}, we surveyed an 18$^{\prime}\times$18$^{\prime}$ region centered on [$l$,$b$] $=$ [346.99, $-$0.41] (southern-most region indicated in Figure\,\ref{fig:CO_wHESS}), and found dense gas associated with molecular cores A, C and D (see Figure\,\ref{fig:CS_-10}). In this study we extend the dense gas survey with another two mapped regions, to encompass Core\,D and the northern peak in TeV $\gamma$-ray emission, and several bright keV X-ray emission features. CS(1-0) emission can probe the dense gas towards \RXJ , which, as highlighted earlier, may be important in a hadronic scenario. The Mopra radio telescope spectrometer is capable of simultaneously recording CS(1-0) emission, while observing the shock-tracing transition SiO(1-0). This allows us to further test the $\sim$1\,kpc RX\,J1713.7 $-$3946 kinematic distance solution through attempting to trace the SNR shock, while undertaking our survey of dense gas. The shock-tracing molecule, SiO, has been observed towards shocked gas associated with other (albeit older) SNRs (eg. \citeauthor{Ziurys:1989}, \citeyear{Ziurys:1989} and \citeauthor{Nicholas7mm:2012}, \citeyear{Nicholas7mm:2012}), so may possibly be present towards gas associated with \RXJ .

We also note the usefulness of other molecular lines simultaneously observed at 7\,mm with the Mopra telescope. These include emission from isotopologues of CS, which are useful for probing optically-thick regions, the vibrationally-excited modes of the SiO(1-0) rotational transition (v=1,2,3), which can sometimes be emitted in association with star formation or evolved stars, and the CH$_3$OH(7(0)-6(1)) transition which can highlight warm regions where CH$_3$OH (methanol) is evaporated from dust grains.

%In an effort to continually test the favoured \RXJ\ distance solution, millimetre indicators of shock-interaction could be of relevance. The shock-tracing molecule, SiO, has been observed towards shocked gas associated with other (albeit older) SNRs (eg. \citeauthor{Ziurys:1989}, \citeyear{Ziurys:1989} and \citeauthor{Nicholas7mm:2012}, \citeyear{Nicholas7mm:2012}), so may possibly be present towards gas associated with \RXJ .

%The Mopra radio telescope spectrometer is capable of simultaneously recording the CS(1-0) and SiO(1-0) transitions, allowing both an investigation of dense gas towards \RXJ\ (via CS), while searching for SiO to provide a kinematic solution for shock-gas interactions, like those expected towards SNRs that possibly accelerate (or have previously accelerated) CRs via the first-order fermi process.

\section{Observations}
In April, 2011, we recorded and co-added 6 Mopra OTF (on the fly) 19$^{\prime}\times$19$^{\prime}$ area maps centered on [$l,b$] $=$[347.36,$-$0.09], and 5 Mopra OTF 16$^{\prime}\times$16$^{\prime}$ area maps centered on [$l,b$]=[347.07,$-$0.11], to produce a data cube with 2 spatial (long/lat) and 1 spectral (velocity) dimension at 7\,mm wavelengths. These maps were added to our map of an 18$^{\prime}\times$18$^{\prime}$ region centered on [$l,b$]=[346.991,$-$0.408], from our previous investigation \citep{Maxted:2012}.

For all our Mopra mapping data, the cycle time is 2.0\,s and the spacing between scan rows is 26$^{\prime\prime}$. The velocity resolution of the 7\,mm zoom-mode data is $\sim$0.2\,kms$^{-1}$. The beam FWHM and the pointing accuracy of Mopra at 7\,mm are 59$\pm$2$^{\prime\prime}$ and $\sim$6$^{\prime\prime}$, respectively. The Mopra spectrometer, MOPS, was employed and is capable of recording sixteen tunable, 4096-channel (137.5\,MHz) bands simultaneously when in `zoom' mode, as used here. A list of measured frequency bands, targeted molecular transitions and achieved $\Trms$ levels are shown in Table\,\ref{Table:bands}.

OTF-mapping and deep ON-OFF switched pointing data were reduced and analysed using the ATNF analysis programs, \textsc{Livedata}, \textsc{Gridzilla}, \textsc{Kvis}, \textsc{Miriad} and \textsc{ASAP}\footnote{See http://www.atnf.csiro.au/computing/software/}. We assumed the beam efficiencies presented in \citet{Urquhart:2010} to convert antenna intensity into main-beam intensity (0.43 and 0.53 for point source CS(1-0) and SiO(1-0) emission, respectively).

In addition to mapping, two deep ON-OFF switched pointings were performed in response to a detection of SiO(J=1-0, v=2) emission (see \S\ref{sec:SiO_2011}). The first of these two pointings was interrupted after achieving a signal-noise ratio similar to mapping data, so a second pointing with a longer exposure-time was performed 12 days later.

\section{Spectral Line Analysis}
We use the CS(1-0) analysis outlined in \citet{Maxted:2012}, and we briefly summarise it here. Gaussian functions were first fitted to all CS(1-0) emission lines using a $\chi ^2$ minimisation method. Generally one can estimate CS(1-0) optical depth by comparing the CS(1-0) intensity to the C$^{34}$S(1-0) intensity, while assuming an abundance ratio, but of the new positions that exhibited CS(1-0) emission, none had corresponding C$^{34}$S(1-0) detections. We constrain optical depth by placing an optically thin (optical depth, $\tau\rightarrow$0) lower-limit and a conservative upper-limit derived using the C$^{34}$S(1-0)-band \Trms\ as an upper limit on C$^{34}$S(1-0) intensity, and assuming a [CS]/[C$34$S] ratio of 22.5.

The calculated optical depth range is used to constrain the CS(J=1) column density (Equation\,9, \citeauthor{Goldsmith:1999}, \citeyear{Goldsmith:1999}), and with an assumption of Local Thermodynamic Equilibrium (LTE) at a temperature of 10\,K, we converted this to total CS column density, $N_{CS}$, (CS column density $\sim$ 3.5$\times$ CS J=1 col. dens.). We note that a 50\% error in the assumed temperature of 10\,K would result in a 20-30\% systematic uncertainty in the column density using this LTE method \citep{Maxted:2013}.

Assuming an abundance of CS with respect to molecular hydrogen, [CS]/[H$_2$]$\sim$10$^{-9}$ \citep{Frerking:1980}, a hydrogen column density is estimated, allowing the estimation of H$_2$ mass. We use no beam-filling correction and assume spherical clumps of a size equal to the 7\,mm beam FWHM (radius$\sim$30$^{\prime\prime}$). Where significant CS(1-0) emission extended beyond an approximate beam-area, sources were divided into segments (east and west in this study) for parameter calculations, and an additional large-scale region was used for an over all mass estimation of the full extent of the CS emission.

\section{Results and Discussion}
This investigation mapped regions indicated in Figures \ref{fig:CO_wHESS} and \ref{fig:CO_xmm} in the CS(1-0) transition towards the RX\,J1713.7$-$3946 SNR. Significant detections at a velocity consistent with CO(1-0,2-1)-traced cores that are believed to be associated with \RXJ\ (line-of-sight velocity, \vlsr$\sim -10$\,km\,s$^{-1}$ according to Moriguchi \etal, 2005; Fukui \etal, 2008) were recorded. In addition to this, we found gas consistent with the background Norma arm (\vlsr$\sim -70$\,km\,s$^{-1}$) and the 3\,kpc-expanding arm (\vlsr$\sim -120$\,km\,s$^{-1}$). For the latter arm, we also note the detection of the CH$_3$OH(7-6) transition.

In addition to our CS and CH$_3$OH detections, transient detections of the v=1 and 2 vibrational modes of the SiO(1-0) rotational transition were present towards one location (\vlsr$\sim -5$\,km\,s$^{-1}$).

\subsection{CS(1-0) Emission}\label{sec:CS}
The CS(1-0) transition has a critical density for emission of $\sim$10$^5$\,cm$^{-3}$ (at temperature $\sim$10\,K) and is ideal for probing the deep, inner regions of molecular clouds. The dense gas traced in this study is displayed in Figures \ref{fig:CS_-10}, \ref{fig:CS_-70} and \ref{fig:CS_-120} and are discussed in the following sections. Table\,\ref{tab:Gas_Params} is a compilation of spectral line fit parameters and gas parameters, including line-of-sight velocity, line-width, optical depth, H$_2$ column density and mass. These values are used as a basis for the discussion of dense gas towards \RXJ .

\subsubsection{CS(1-0) Emission Between \\\vlsr$\sim -$12.5 and $-$7.5\,km\,s$^{-1}$}\label{sec:ResDis}
\begin{figure*}[!h]
\centering
\includegraphics[width=0.99\textwidth]{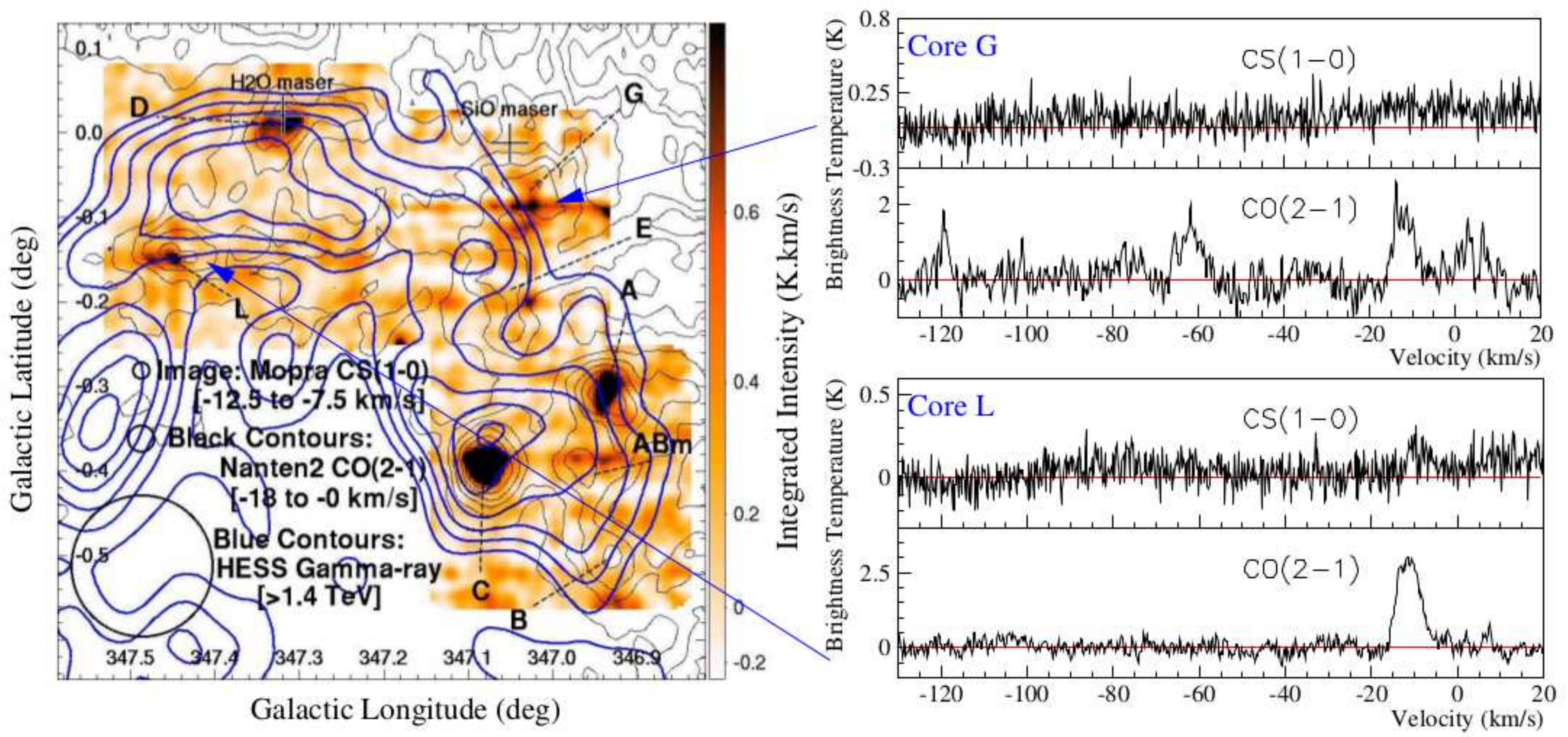}
\caption{Left Panel: Colour image of integrated CS(1-0) emission (\vlsr$=-$12.5~to~$-$7.5\,km\,s$^{-1}$) from Mopra overlaid with black contours of CO(2-1) emission (\vlsr$=-$18~to~0\,km\,s$^{-1}$) from Nanten2 \citep{Fukui:2008}, as well as solid blue contours of HESS $>$1.4\,TeV excess emission (same levels as for Figure\,\ref{fig:CO_wHESS}). Core names and the position of masers are indicated. CO(2-1) emission contour-levels are 5, 10, 15, 20, 25, 30, 35 and 40\,K\,km\,s$^{-1}$. Right Panel: We also show spectral profiles of molecular emission towards locations of interest (indicated).\label{fig:CS_-10}}
\end{figure*}

Figure\,\ref{fig:CS_-10} is an image of CS(1-0) emission between \vlsr$=-$12.5 and $-$7.5\,km\,s$^{-1}$, which features detections corresponding to cores believed to be associated with the RX\,J1713.7$-$3946 SNR.

The dense components of cores A, B, C and D were previously studied in detail by us \citep{Maxted:2012}, revealing multiple detections of 6 species (and several isotopologues), including CS(1-0) emission towards Cores A, C and D. This allowed the mass of the dense molecular gas to be estimated and these are re-stated (along with our new results) in Table\,\ref{tab:Gas_Params}.

In our new survey (this work), we revealed CS(1-0) emission towards Cores G and L, with peak intensities of $\sim$1.5\,\Trms. After integration (between velocities, $\vlsr -$12.5 and $-$7.5\,km\,s$^{-1}$), Cores G and L were detected at a pre-trial level of $\sim$4-4.5$\sigma$. Since the CS(1-0) emission intensities in these 2 cores are very low, the resultant CS(1-0) optical depth estimates are poorly-constrained. However, given that the CS(1-0) intensities of Cores G and L are relatively low compared to those of Cores A, C and D (which have LTE masses in the range 12\,-\,120\,M$_{\odot}$), we favour the optically thin solutions. It follows that dense molecular gas masses of Cores G and L are on the order of $\sim$1\,M$_{\odot}$, and do not represent a significant proportion of mass in the \RXJ\ field at \vlsr$\sim -$10\,km\,s$^{-1}$ ($<$0.1\%), therefore they probably do not play an important role in the dynamics of the SNR-shock or the production of $\gamma$-ray emission.

Indeed, it appears that the total CS(1-0)-derived mass of all the surveyed cores represents only a fraction of the total gas mass in the region. The total CO(1-0)-derived mass of $\sim$1$\times$10$^{4}$\,M$_{\odot}$ towards \RXJ\ \citep{Fukui:2012} is significantly larger than the total CS(1-0)-derived mass (dense core) components of mass 80\,-\,200\,M$_{\odot}$ in our surveyed region (a subset containing the densest molecular gas of the CO-traced region). % smaller region surveyed in 7\,mm (as opposed to CO surveys), but because we surveyed the region containing the majority of mass of molecular gas, 

Previous authors have noted a small-scale anticorrelation between CO emission peaks at $\sim$1\,kpc (\vlsr$\sim -$18-0\,km\,s$^{-1}$) and X-ray emission (see \S\ref{sec:intro}), suggesting a SN shock interaction with gas. Indeed, two X-ray peaks lie on the outskirts of the CS(1-0) (and CO(4-3,7-6), \citeauthor{Sano:2010}, \citeyear{Sano:2010}) emission corresponding to Core\,C (see Figure\,\ref{fig:CS_xmm}) and a similar X-ray peak lies coincident with Core\,D. The same cannot be said for Cores A, G and L, but, although the densest regions of Cores G and A sit outside the \RXJ\ X-ray boundary, X-ray flux peaks do seem to correspond to regions directly adjacent to the CS(1-0) boundaries. This possibly indicates a shock-compression of the Core G and A gas, however we have previously noted that these features may alternatively be consistent with photoelectric absorption of X-rays emitted behind the molecular cores \citep{Maxted:2012}.

Previous authors (e.g. \citeauthor{Moriguchi:2005}, \citeyear{Moriguchi:2005}) have noted indicators of star formation towards RX\,J1713.7 $-$3946, and new data from HOPS (H$_2$O Plane Survey, \citeauthor{Walsh:2011}, \citeyear{Walsh:2011}) supports this picture. On examination of the HOPS catalogue, we found narrow (\fwhm$\sim$1.1\,km\,s$^{-1}$) 12\,mm H$_2$O maser emission at a velocity and position consistent with Core\,D (G347.32$+$0.2, \vlsr$\sim -$8.5\,km\,s$^{-1}$). The location of this maser is indicated in Figures \ref{fig:CS_-10} and \ref{fig:CS_xmm}.

\subparagraph{Gas as CR Target Material:}
The gas between \vlsr$\sim -$12.5 and $-$7.5\,km\,s$^{-1}$ may be acting as CR target material in a hadronic gamma-ray emission scenario for \RXJ, so we investigate the effect of the mass of CS(1-0)-traced gas.

\citet{Aharonian:1991} derived a relation to predict the hadronic flux of $\gamma$-rays above a given energy from the mass of CR-target material, assuming an $E^{-1.6}$ integral power law spectrum. We can calculate the expected $\gamma$-ray flux above energy $E_{\gamma}$,
\begin{equation}
\label{equ:Aharonian}
F\left(\geq E_{\gamma} \right) = 2.85\times 10^{-13} E_{TeV}^{-1.6} \left( \frac{M_5}{d_{kpc} ^2 }\right) k_{CR}~~~~\textrm{cm$^{-2}$s$^{-1}$}
\end{equation} where $M_5$ is the gas mass in units of 10$^5$\,M$_{\odot}$, $d_{kpc}$ is the distance in units of kpc, $k_{CR}$ is the CR enhancement factor above that observed at Earth, $E_{TeV}$ is the lower $\gamma$-ray energy in units of TeV and $E_{\gamma}=$1\,TeV.

Given that in a hadronic scenario $\gamma$-ray flux is proportional to mass, one can assume that the dense CS(1-0)-traced component ($\sim$few percent the total mass) may only account for $\sim$few percent of the total $\gamma$-ray flux. In a lepton-dominated scenario, a hadronic component would comprise an even smaller proportion of the total flux (eg. \citeauthor{Zirakashvili:2010}, \citeyear{Zirakashvili:2010}).%, making the hadronic $\gamma$-ray flux from CS(1-0)-traced regions even less significant.

\begin{figure*}%[!h]
\centering
\includegraphics[width=0.99\textwidth]{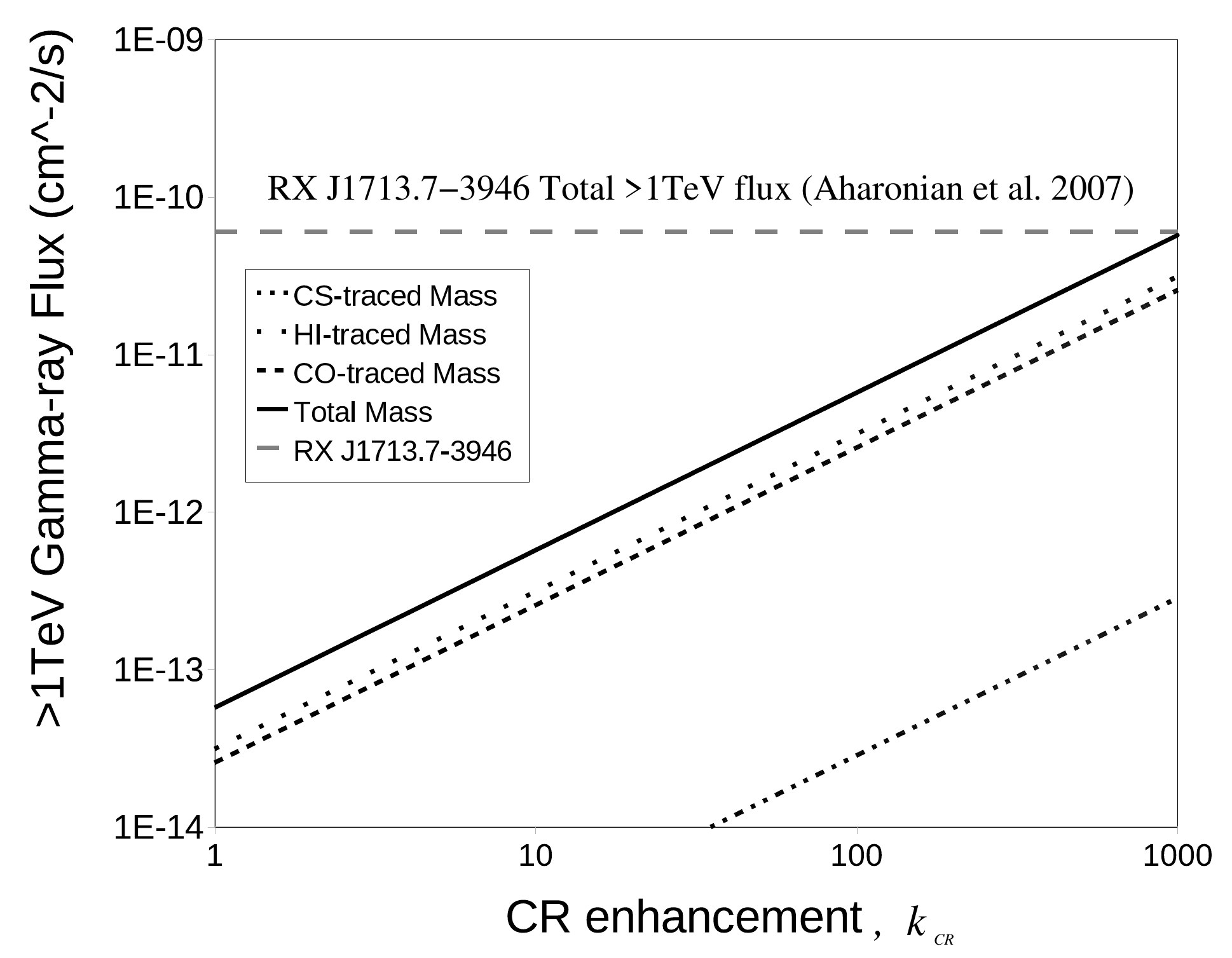}
\caption{The predicted hadronic $\gamma$-ray flux as a function of CR-enhancement for gas components possibly associated with \RXJ\ (\vlsr$\sim -$10\,km\,s$^{-1}$). The contributions from HI and CO-traced components towards the \RXJ\ region and CS(1-0) emission from 5 cores are included (this work). The total predicted hadronic $\gamma$-ray flux remains less than the measured \RXJ $\gamma$-ray flux for CR enhancement values less than 1000. A distance of 1\,kpc was assumed.\label{fig:CRenhanceRXJ}}
\end{figure*}

Figure\,\ref{fig:CRenhanceRXJ} is a graph of the predicted hadronic $\gamma$-ray flux above 1\,TeV from various gas components (traced using HI, CO and CS) assumed to be acting as target material for CRs accelerated in the \RXJ\ shock at a distance of 1\,kpc (using Equation\,\ref{equ:Aharonian}). It is predicted that the CR enhancement would have to be as high as $k_{CR} \sim$1\,000 for the \RXJ\ gamma-ray flux above 1\,TeV, of $\sim$6$\times$10$^{-11}$\,cm$^{-2}$\,s$^{-1}$, to be entirely from hadronic processes. In such a scenario, the dense gas components (as traced by CS(1-0) emission) would contribute to a gamma-ray flux of $\sim$3$\times$10$^{-13}$\,cm$^{-2}$\,s$^{-1}$. This flux may be detectable with the current HESS sensitivity, but the HESS beam FWHM is too large to resolve molecular cores at a scale of $\sim$1\,$^{\prime}$. Future experiments such as the CTA gamma-ray telescope \citep{CTA:2010} may reach the arcminute resolution required to do this.

A CR enhancement factor of $k_{CR} \sim$1\,000 towards \RXJ\ is plausible, supported by work by \citet{Aharonian:1996}. Figure\,1b of this paper displays cosmic ray enhancement factors of the order $k_{CR} \sim$10$^3$ at a distance of 10\,pc from a SNR of age 10$^3$\,yr (similar to the \RXJ\ age). The source spectral index in these simulations was 2.2 and the CR diffusion coefficient was 10$^{26}$\,cm$^2$\,s$^{-1}$ (slow diffusion like that seen towards, W28, e.g. \citeauthor{Gabici:2010}, \citeyear{Gabici:2010}).

We estimate the total cumulative kinetic energy of CRs above an energy of 1\,GeV (assuming a pure E$^{-2.6}$ power law down to 1\,GeV) in the \RXJ\ region, to be,
\begin{align}
\label{equ:EnCRs}
E_{CR\,tot} &=E_p n_{CR}\,dV\,dE_P \nonumber \\
&\sim \frac{4}{3}\pi r_{SNR}^3 \int _{1\,\textrm{GeV}} ^{\infty} E_P n _{CR}\,dE_P \nonumber \\ 
&\sim (1.2\times 10^{59}\,\textrm{cm}^3) (5.6 \times 10 ^{-11}\,\textrm{GeV\,cm}^{-3})k_{CR}\nonumber \\ 
&\sim k_{CR}\times 10^{46}\textrm{\,erg}
%&\sim \frac{4}{3}\pi r_{SNR}^3 \int _{1\,\textrm{TeV}} ^{\infty} E_P n _{CR}\,dE_P \nonumber \\ 
%&\sim (1.2\times 10^{59}\,\textrm{cm}^3) (8.8 \times 10 ^{-13}\,\textrm{GeV\,cm}^{-3})k_{CR}\nonumber \\ 
%&\sim 2k_{CR}\times 10^{44}\textrm{\,erg}
\end{align}
% $E_{CR\,tot}=E_p n_{CR}\,dV\,dE \sim(4/3)\pi r_{SNR}^3 k \int _{1} ^{\infty} E_P n _{CR}\,dE \sim 4k\times10^{43}$\,erg,
where $V$ is the SNR volume, $E_P$ is CR energy, $n_{CR}$ is the cosmic ray density and $r_{SNR}$ is the radius of a spherical region of CR enhancement, $k_{CR}$ (assumed to be 10\,pc, consistent with the RX\,J1713.7 $-$3946 radius). Assuming a SN blast kinetic energy of 10$^{51}$\,erg, a CR enhancement of 1\,000 would correspond to $\sim$1\% of the SN blast kinetic energy being injected into $>$1\,GeV protons, so energetically, a hadronic scenario is consistent.
% Assuming a SN blast kinetic energy of 10$^{51}$\,erg, a CR enhancement of 1\,000 would correspond to $\sim$0.02\% of the SN blast kinetic energy being injected into $>$1\,TeV protons. Extrapolating the CR spectrum down to 1\,GeV implies that about $\sim$1\% of the SN blast kinetic energy has gone into Galactic CR acceleration, so energetically, a hadronic scenario is consistent.

The presented cosmic ray enhancement factor values were calculated under the assumption that all the gas in the region is taken into account, well-traced by the latest HI, CO and CS studies \citep{Fukui:2012,Maxted:2012b}, which may not be entirely valid. A so-called `dark' component of gas, where carbon is in atomic/ionic form (not in molecules such as CO or CS) and hydrogen is in molecular form (therefore does not emit the atomic H 21\,cm line) exists and isn't taken into account by the tracers considered in studies of \RXJ . \citet{Wolfire:2010} predicted that this dark component may comprise $\sim$ 30\% of the total mass of an average cloud but could be even higher in regions with enhanced CR ionisation rates, as is likely around RX\,J1713.7$-$3946.

Our method also required the assumption of an $E^{-1.6}$ integral power law CR spectrum, which is not supported by studies of the RX\,J1713.7$-$3946 high energy photon spectrum. The RX\,J1713.7$-$3946 $\gamma$-ray spectrum is consistent with an $E^{-1.3}$ CR integral spectrum, reflective of a pure power law $\chi^2$-minimisation fit to the $\gamma$-ray spectrum \citep{Aharonian:2007}. 
Using the fact that the total energy injected in CRs bewteen 1\,GeV and 1\,PeV is expected to be approximately fixed ($\sim$10$^{50}$\,erg), we can pivot the assumed CR energy spectrum about an energy of $\sim$1\,TeV. 
%If we assume that the total energy injected into CR protons between 1\,GeV and 1\,PeV is conserved, regardless of the spectral slope, we can pivot the assumed CR energy spectrum about an energy of $\sim$1\,TeV.
This would correspond to a pivot energy of $\sim$0.1\,TeV in the resultant hadronic gamma-ray spectrum. It follows that Equation\,\ref{equ:Aharonian} can be scaled to become $F\left(\geq E_{\gamma} \right) \sim 4.6\times 10^{-13} E_{TeV}^{-1.3} \left( M_5/d_{kpc} ^2 \right) k_{CR}$.
The resultant CR enhancement factors (for CR energy $\geq$1\,TeV) presented in this paper would thus reduce by $\sim$30\%.
%The resultant CR enhancement factors (for CR energy $\geq$1\,TeV) presented in this paper may thus be overestimated by $\sim$40\%.}

It's clear that in a hadronic scenario, the atomic and molecular mass traced by HI and CO would be the dominant contributor to $\gamma$-ray emission. Despite this, tell-tale signs of hadronic emission might feasibly be detected in the form of small-scale hardening and localisation of high energy spectra correlated with dense gas components. Such phenomena would result from energy-dependent CR diffusion into dense gas in turbulent regions that experience a suppressed diffusion level. This might act to hinder the transport of lower energy CRs more than higher energy CRs and result in a harder than average spectrum (eg. \citeauthor{Gabici:2009}, \citeyear{Gabici:2009}; \citeauthor{Casanova:2010}, \citeyear{Casanova:2010}; \citeauthor{Fukui:2012}, \citeyear{Fukui:2012}; \citeauthor{Maxted:2012}, \citeyear{Maxted:2012}).

In a previous paper \citep{Maxted:2012}, we estimated the level of penetration of CRs of a range of energies (10\,GeV to 10$^{2.5}$\,TeV) into a homogenous molecular region (average density 300\,cm$^{-3}$) of radius 0.62\,pc (similar to Core\,C) within a time equal to the \RXJ\ age ($\sim$1\,600\,yr). We found that for decreasing values for the diffusion-suppression coefficient, $\chi$ (see Equation\,\ref{equ:Diffcoeff}), the level of CR penetration into molecular cores may plausibly be decreased, particularly at lower energies.

For a plausible guide on the level of CR penetration into dense gas, we used the 3D random walk distance, $d=\sqrt{6\,D\,t}$, where $t$ is time and the diffusion coefficient, $D$, was parametrised as,
\begin{equation}
\label{equ:Diffcoeff}
D(E_P,B)=\chi D_0 \left( \frac{E_P/\textrm{\small{GeV}}}{B/3\,\textrm{\small{$\mu$G}}}\right)^{0.5}~~~\textrm{\small{[cm$^2$\,s$^{-1}$]}},
\end{equation} where $D_0$ is the galactic diffusion coefficient, assumed to be 3$\times$10$^{27}$\,cm$^2$\,s$^{-1}$ to fit CR observations \citep{Berezinskii:1990}, and $\chi$ is the diffusion suppression coefficient (assumed to be $<$1 inside the core, 1 outside), a parameter invoked to account for possible deviations of the average galactic diffusion coefficient inside molecular clouds \citep{Berezinskii:1990,Gabici:2007}. $E_P$ was the proton energy and $B$ was the magnetic field, calculated to be consistent with \citet{Crutcher:1999}.

We found that for $\chi=$10$^{-4}$ and $\chi=$10$^{-5}$, CR protons with energy $<$10\,TeV were somewhat excluded from the central $\sim$0.2 and $\sim$0.5\,pc radius, respectively. The total level of exclusion, however, is difficult to confidently constrain from such a simplistic approach, so we are currently working on diffusion models that numerically solve the diffusion equation to accurately predict the CR distribution, hence the change in CR enhancement inside dense gas (details to be presented in a future paper). It is expected that diffusion -dependent modeling will yield CR enhancement factors that are functions of not only CR energy, but gas density, i.e. the CR enhancement factor, $k_{CR}$, might decrease towards dense clumps, the effect lessening for increasing energy. The result would be harder-than-average hadronic $\gamma$-ray components correlated with dense gas. The prediction of features in $\gamma$-ray spectra will result from future diffusion analyses.

\subsubsection{CS(1-0) Emission at Other Velocities}
\begin{figure*}[!h]
\centering
\includegraphics[width=0.99\textwidth]{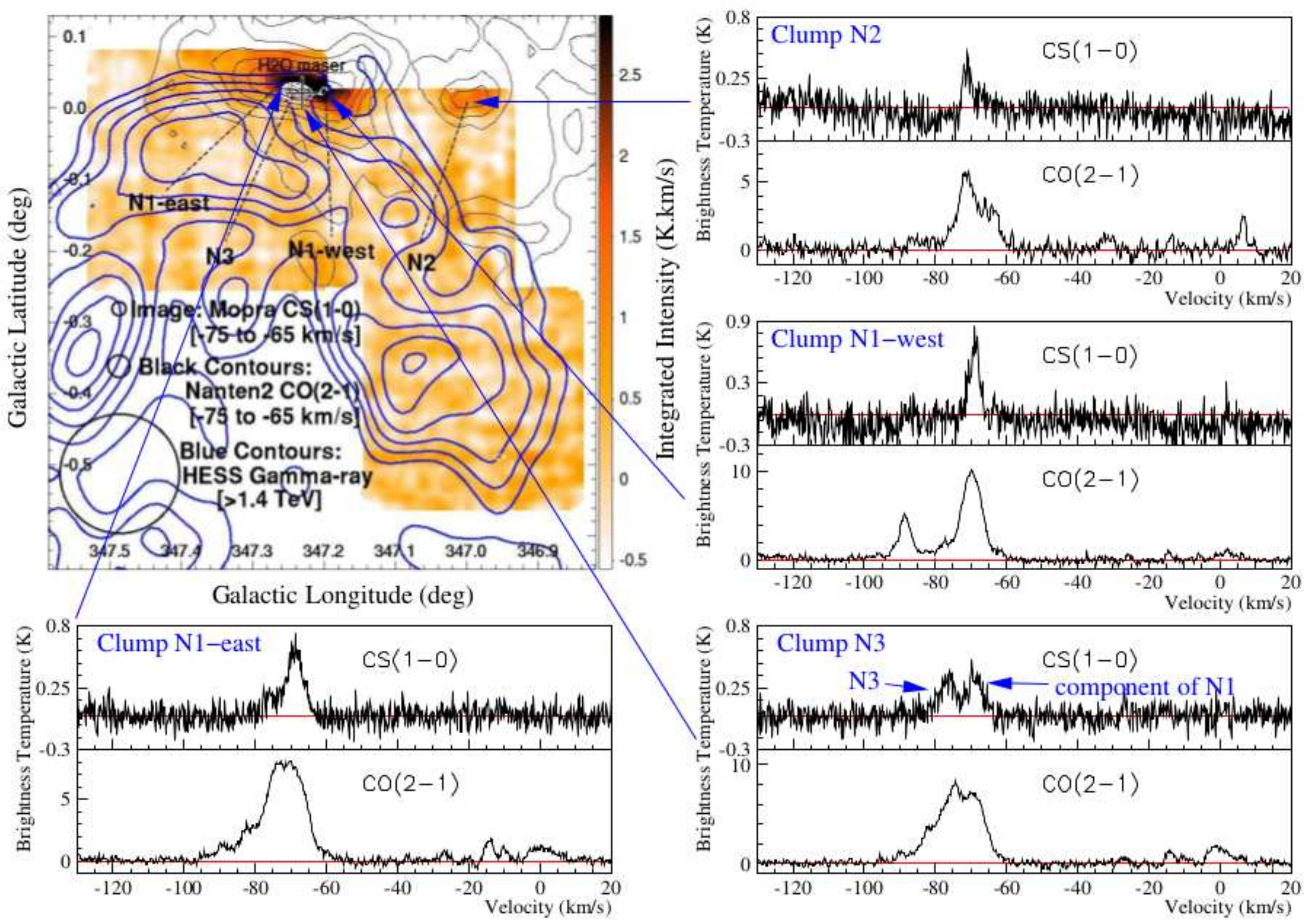}
\caption{Top Left Panel: Colour image of integrated CS(1-0) emission (\vlsr$=-$75~to~$-$65\,km\,s$^{-1}$) from Mopra overlaid with black contours of CO(2-1) emission (\vlsr$=-$75~to~$-$65\,km\,s$^{-1}$) from Nanten2 \citep{Fukui:2008}, as well as solid blue contours of HESS $>$1.4\,TeV excess emission (same levels as for Figure\,\ref{fig:CO_wHESS}). CO(2-1) emission contour-levels are 10, 20, 30, 40, 50 and 60\,K\,km\,s$^{-1}$. White contours indicate Clump\,N3 Nanten2 integrated CS(1-0) emission (\vlsr$=-$80~to~$-$72.5\,km\,s$^{-1}$) with contour-levels of 0.8, 1.0, 1.2, 1.4 and 1.6\,K\,km\,s$^{-1}$. The position of an H$_2$O maser is indicated. Right/Bottom Panels: We also show spectral profiles of molecular emission towards locations of interest (indicated).\label{fig:CS_-70}}
\end{figure*}

\begin{figure*}[!h]
\centering
\includegraphics[width=0.99\textwidth]{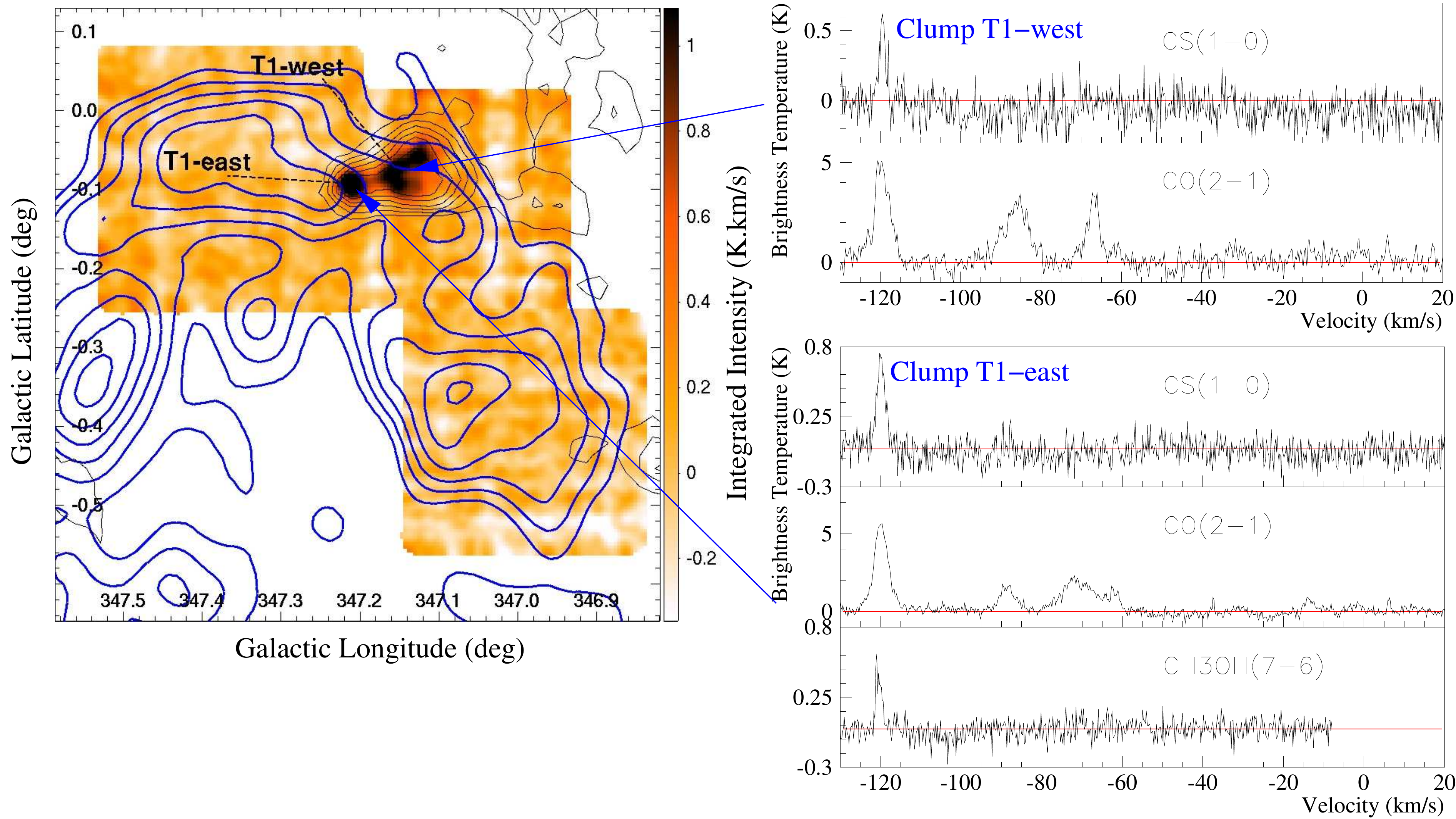}
\caption{Top Left Panel: Colour image of integrated CS(1-0) emission (\vlsr$=-$122~to~$-$115\,km\,s$^{-1}$) from Mopra overlaid with black contours of CO(2-1) emission (\vlsr$=-$122~to~$-$115\,km\,s$^{-1}$) from Nanten2 \citep{Fukui:2008}, as well as solid blue contours of HESS $>$1.4\,TeV excess emission (same levels as for Figure\,\ref{fig:CO_wHESS}). CO(2-1) emission contour-levels are 2, 4, 6, 8, 10 and 12\,K\,km\,s$^{-1}$. Right Panels: We also show spectral profiles of molecular emission towards locations of interest (indicated).\label{fig:CS_-120}}
\end{figure*}

In addition to the gas believed to be in association with \RXJ\ (see \S \ref{sec:ResDis}), dense gas components corresponding to background gas clouds were also detected. Prominent CS(1-0) emission from molecular clouds in the Norma arm and the 3\,kpc-expanding arm are displayed in Figures \ref{fig:CS_-70} and \ref{fig:CS_-120}, respectively.

%The \RXJ\ distance hasn't always been clear and there is still some controversy to this day.
On the question of the \RXJ\ distance (currently thought to be $\sim$1\,kpc), some previous molecular gas surveys favoured an association with gas that featured broad CO emission components at a distance of $\sim$6.3\,kpc (\vlsr$\sim -$90\,km\,s$^{-1}$). It was argued that an enhanced CO(2-1)/CO(1-0) intensity ratio is characteristic of shocked gas \citep{Slane:1999,Butt:2001}. The optically subluminous nature of the RX J1713.7$-$3946 progenitor event (see \S\ref{sec:intro}) may also be considered to support a larger distance \citep{Fessen:2012}. For completeness, we consider these CS(1-0) detections in our analyses.

\subparagraph{The Norma Arm\\(\vlsr$\sim -75$ to $-65$\,km\,s$^{-1}$):}
At a distance of $\sim$6\,kpc, in the northernmost regions of the surveyed field (Figure\,\ref{fig:CS_-70}), are CS(1-0)-traced dense gas components. These components, labelled here as N1, N2 and N3 are within the Norma arm and are coincident with peaks in CO(2-1) emission. Due to the extended nature of N1, this region had previously been scrutinised in two sections N1-east and N1-west, with the calculated masses being of the order of a few$\times$10$^3$\,M$_{\odot}$ for both sides \citep{Maxted:2012b}. Further scrutiny of spectral data resulted in the discovery of another coincident dense core component, N3. Clump N3, which is comparable in mass to both N1-east and west ($\sim$10$^3$\,M$_{\odot}$), is apparent as one peak of the double-peaked profile in the bottom-right spectra of Figure\,\ref{fig:CS_-70}, (whereas the other peak is an edge of N1) and is shown as white contours in the image. Clump N2, in the north-west of the image in Figure\,\ref{fig:CS_-70} has a mass on the order of $\sim$10$^3$\,M$_{\odot}$.

Clumps N1 and N3 are coincident with the boundary of the \RXJ\ HESS TeV $\gamma$-ray excess, so might feasibly be CR target material associated with \RXJ\ if \RXJ\ is associated with the Norma arm gas (rather than the currently-favoured \vlsr$\sim -$10\,km\,s$^{-1}$/1\,kpc-distance gas), and a hadronic scenario is applicable. In contrast, clump N2 lies outside of the \RXJ\ TeV $\gamma$-ray boundary, so cannot contribute to the total hadronic $\gamma$-ray flux for any distance solution.

\begin{figure*}[!h]
\centering
\includegraphics[width=0.48\textwidth]{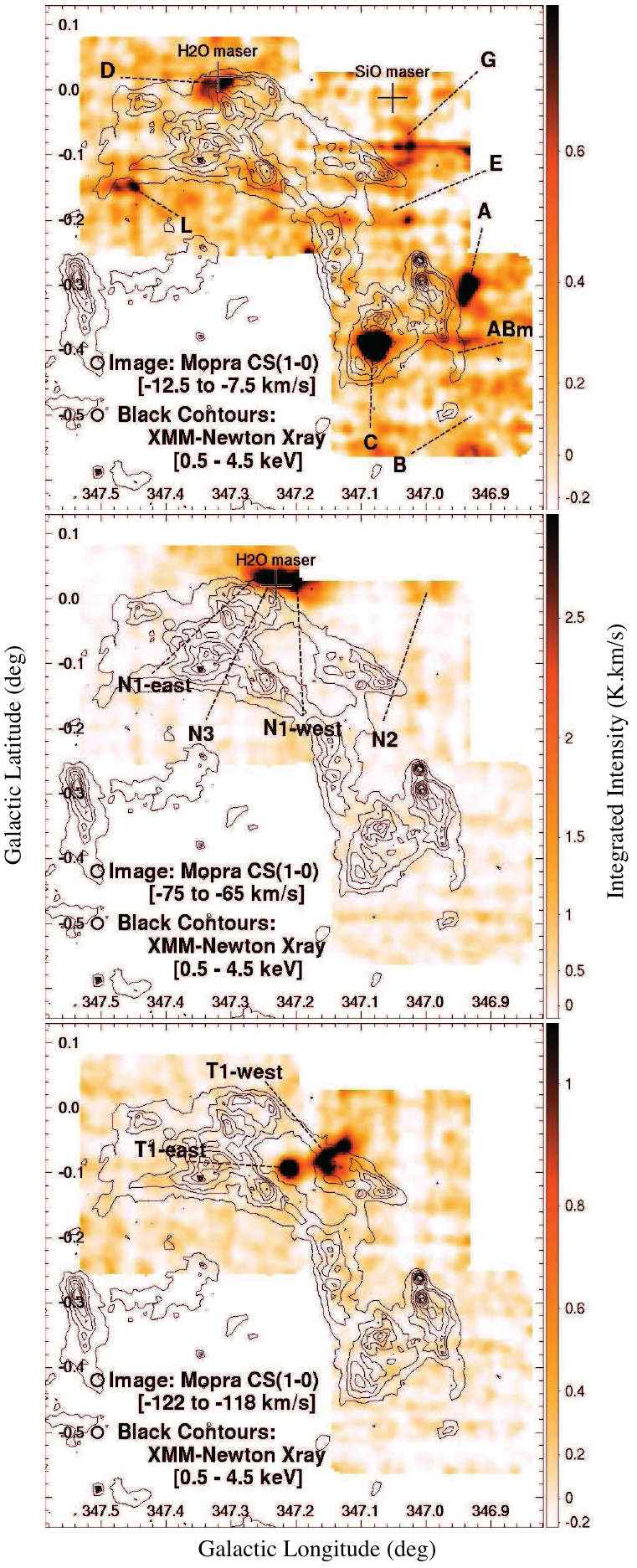}
\caption{Colour image of integrated CS(1-0) emission, shown for three different ranges indicated in each panel, overlaid with XMM-Newton 0.5-4.5\,keV X-ray contours from \citet{Acero:2009}. X-ray contours span 0.003-0.015\,cm$^{-2}$\,s$^{-1}$ in increments of 0.003\,cm$^{-2}$\,s$^{-1}$. Core names and the position of masers are indicated. The Mopra 7\,mm and XMM-Newton beam FWHM are represented by the small circles next to the text on the picture.\label{fig:CS_xmm}}
\end{figure*}

The dense regions of Clumps N1 and N3 lie coincident with the northern boundary of the keV X-ray emission (see Figure\,\ref{fig:CS_xmm}), which may be consistent with an RX\,J1713 $-$3946 shock-interaction with Norma-arm gas in the north, but this correspondence may be less convincing than the north-to-east shock-correspondence of the gas at \vlsr$\sim -$10\,km\,s$^{-1}$ (see \S\ref{sec:ResDis}). CS(1-0) spectral lines from Clumps N1 and N3 are quite broad (\vlsr$\sim$4.5-8\,km\,s$^{-1}$). Such broadening is consistent with a shock-gas interaction, but also consistent with multiple components of warm, turbulent gas. The latter scenario is supported by 8 and 24\,$\mu$m data (see Figure\,\ref{fig:RXJ_infrared}), which shows several peaks coincident with Clumps N1 and N3. Indeed, we note that narrow (\fwhm$\sim$1.2\,km\,s$^{-1}$) 12\,mm H$_2$O maser emission (G347.23$+$0.02, \vlsr$\sim -$77.8\,km\,s$^{-1}$, \citeauthor{Walsh:2011}, \citeyear{Walsh:2011}) is indicative of star formation within the N1 or N3 regions (see Figures \ref{fig:CS_-10} and \ref{fig:CS_xmm}), the velocity of the maser being closest to that of the clump N3 (the N3 component has velocity, \vlsr$\sim -$76.6$\pm$0.3\,km\,s$^{-1}$).

\subparagraph{The 3\,kpc-Expanding Arm\\(\vlsr $\sim -122$ to $-118$\,km\,s$^{-1}$):}
Also coincident with the northern boundary of RX\,J1713 $-$3946 is molecular gas in the 3\,kpc-expanding arm, at a distance of $\sim$6.5\,kpc (Figure\,\ref{fig:CS_-120}). The T1 cloud has a mass on the order of $\sim$10$^3$\,M$_{\odot}$ and contains a distinctive molecular core in the east, T1-east. A hot component in T1-east is evidenced by our detection of CH$_3$OH(7-6) emission, which is emitted after CH$_3$OH is evaporated from dust-grains in hot (Temperature $\sim$100\,K) conditions \citep{vanDishoeck:1998}. This emission is possibly related to the coincident object, S9 of \citet{Churchwell:2006}, which appears as an incomplete ring of 8\,$\mu$m emission (see Figure\,\ref{fig:RXJ_infrared}). The authors argue that such infrared `bubbles' result from hot young stars in massive star formation regions. Our 7\,mm results are consistent with T1-east being a hot star-forming core.

\subparagraph{Background Gas as CR Target Material:}
In Figure\,\ref{fig:CRenhanceBG}, we present the predicted hadronic $\gamma$-ray components from dense gas at distances near $\sim$6\,kpc (T1, N1 and N3), for the scenario where this gas is associated with \RXJ .

The dense gas components in the Norma and 3\,kpc-expanding arms would contribute hadronic $\gamma$-ray emission at a comparable level to that of the dense gas towards the $\sim$1\,kpc region for a given CR enhancement factor, so expected CR enhancement estimates are not likely to help in favouring a specific distance solution. \citet{Moriguchi:2005} systematically compared the angular correlation between the \RXJ\ X-ray emission and CO(1-0) emission at a wide range of line of sight velocities, and concluded that gas in the $\sim$1\,kpc region had better angular correspondence than other line-of-sight gas components (including the Norma and 3\,kpc-expanding arm gas), thus favouring the $\sim$1\,kpc distance for \RXJ . In a hadronic scenario, correlation is expected between gas and $\gamma$-ray emission, so their argument can be extended to include the hadronic production of gamma-rays, in which case gamma-ray correspondance favours the the $\sim$1\,kpc distance. Modelling by \citet{Inoue:2012} found that the large-scale correlation between CO(1-0) and the HESS TeV $\gamma$-ray emission supported the $\sim$1\,kpc gas association, a model also supported by a small-scale gas-keV X-ray anticorrelation.

We note that the coincident dense gas components of these galactic arms, despite being significantly more massive than the $\sim$1\,kpc gas, would produce a flux of $\gamma$-rays ($\sim$4$\times$10$^{-13}$\,cm$^{-2}$\,s$^{-1}$ above 1\,TeV) similar to the dense gas believed to be associated with RX\,J1713.7 $-$3946, due to the greater distance involved.

%T1-west has a similar mass ($\sim$10$^3$\,M$_{\odot}$), but is more extended in nature.

%\subparagraph{Other background Gas}
We further note that CO(2-1) spectra toward N1-west and T1 (Figures \ref{fig:CS_-70} and \ref{fig:CS_-120}, respectively) include a component at \vlsr$\sim -$85\,km\,s$^{-1}$ which doesn't appear to include a detectable dense gas component traced by CS(1-0) emission (above \Trms$\sim$0.16\,K\,ch$^{-1}$). This gas may correspond to `Cloud\,A' \citep{Slane:1999,Butt:2001} or gas traced by \citet{Moriguchi:2005}.

%Prominent CS(1-0) detections corresponding to background gas in the 3\,kpc-expanding arm and the Norma arm are displayed in Figure\,\ref{fig:BGgas}. Dense Norma-arm gas, Clumps N1 and N2 (see Figure\,\ref{fig:BGgas}), lie on the northern border of the surveyed field. 
%CS(1-0) emission is coincident CO(2-1) emission peaks and traces a mass of dense gas of the order 10$^3$-10$^4$\,M$_{\odot}$ (see Table\,\ref{tab:Gas_Params}). Similarly, in the 3\,kpc-expanding arm, Clump\,T1 featured extended CS(1-0) emission coincident with a peak of CO(2-1) intensity, and contains 3-12$\times$10$^3$\,M$_{\odot}$ of dense gas. CH$_3$OH(7-6) emission is detected in Clump\,T1-east (see Figure\,\ref{fig:BGgas}), indicating a localised warm, dense environment.

\subsection{SiO(1-0) Emission}
A detection of vibrational modes (v=1,\,2) of the SiO(1-0) transition towards [$l$,$b$]$\sim$[347.05,$-$0.012] in April, 2011, prompted follow-up observations that were carried out one year later. The position of this detection is indicated in Figure\,\ref{fig:CS_-10}  and spectra are displayed in Figure\,\ref{fig:SiO_Detection}. Spectral characteristics from $\chi ^2$-minimisation fits are presented in Table\,\ref{tab:SiO_Params}.

\subsubsection*{April 2011}\label{sec:SiO_2011}
We measured SiO(J=1-0,v=2) emission that had an intensity greater than the intensity of the lower vibrational mode, v=1, indicating the possible presence of an SiO population-inversion, thus the emission can be considered as a weak SiO maser. The v=1 and v=2 SiO(J=1-0) transitions are at consistent velocities, but if a tentative SiO(J=1-0,v=0) detection is real, there is a discrepency in rest velocity (offset from the v=1 and 2 modes by 1.6\,$\pm$\,0.14\,km\,s$^{-1}$ and 1.4\,$\pm$\,0.1\,km\,s$^{-1}$, respectively). This, in fact, may have been a thermal component, unrelated to the SiO maser, but since this weak signal was only considered (and searched-for) after the detection of the relatively prominent SiO(J=1-0,v=2) emission, we estimate the likelihood that it was noise. A naive post-trial analysis that considers the \vlsr\ range (9\,km\,s$^{-1}$) in the line-fitting process and the fitted line-width (1.5\,km\,s$^{-1}$) leads to an estimate of 6 trials. This decreases the significance from $\sim$1.5\,\Trms\ to an insignificant $\sim$0.65\,\Trms\ (1.5/$\sqrt{6}$). It follows that although a weak thermal SiO emission line may exist towards this location, we dismiss the 2011 SiO(J=1-0,v=0) detection displayed in Table\,\ref{tab:SiO_Params}.

We do not dismiss the 2011 SiO(J=1-0,v=1) detection using a similar argument since the chance probability of fitting a \vlsr\ consistent (within $\sim$0.5\,$\sigma\sim\pm$\,$0.2$\,km\,s$^{-1}$) with the original SiO(J=1-0,v=2) detection is approximately $\sim$5\% (trials in a 9\,km\,s$^{-1}$ \vlsr -range).

\subsubsection*{April 2012}
The deep 7\,mm observations of April 20th, 2012, allowed a $\sim$2.5$\times$ reduction of spectral noise, and the previously tentative (and insignificant after post-trial analyses) SiO(1-0,v=0) line was now absent. Interestingly, the peak intensities of the v=1 and v=2 SiO(J=1-0) emission lines decreased by $\sim$35\% and $\sim$75\%, respectively, while the line-width increased by $\sim$45\% and $\sim$100\%, respectively. 

When we consider the changes in integrated intensity (proportional to the changes in \Tpeak$\times$\fwhm), the variation may seem less dramatic and possibly partly due to a change in the velocity distribution of emitting SiO molecules. The v=1 and v=2 SiO(J=1-0) integrated intensities decrease by $\sim$10\% and $\sim$50\%, respectively. Additionally, the rest velocity was observed to change by $\sim$1.5\,km\,s$^{-1}$.
%Besides the large intensity change, perhaps the most interesting aspect of the transient nature of the SiO maser, is the drastic change in \vlsr\ of $\sim$1.5\,km\,s$^{-1}$.

We note that $\chi ^2$-minimisation gaussian-fits for the April 8th SiO(1-0) observations were divergent, so the full width half maximum was set manually in order to achieve a $\chi ^2$-minimisation fit for the other line parameters. 

\subsubsection*{The SiO Maser Origin}
\begin{figure*}[!h]
\centering
\includegraphics[width=0.99\textwidth]{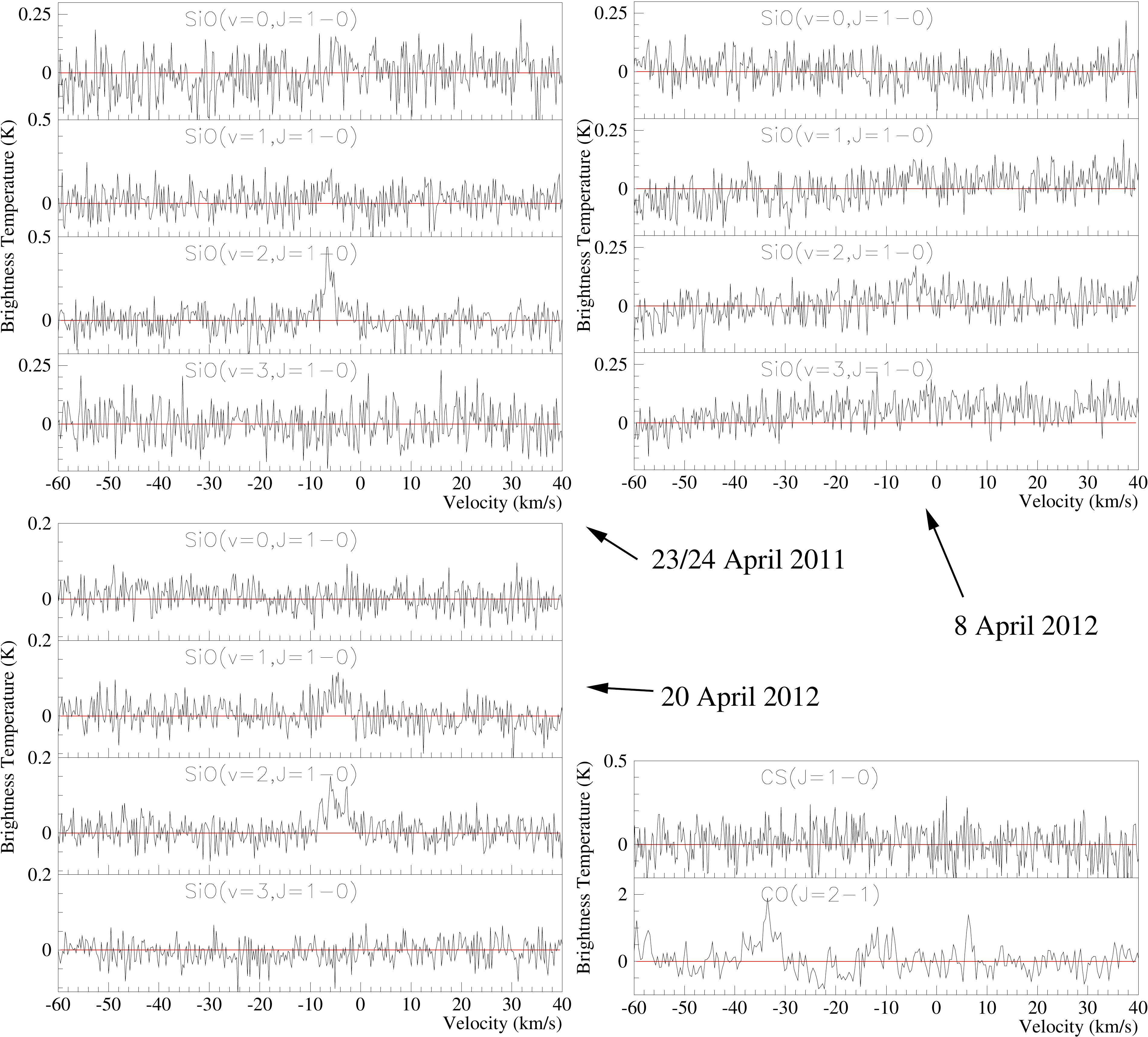}
\caption{Spectra (Mopra and Nanten2) towards the SiO maser discovered at ($l$,$b$)$\sim$(347.05,$-$0.012). Data from the 23rd and 24th of April, 2011 were taken from OTF maps. Data from April, 2012 are from follow-up pointing data. \label{fig:SiO_Detection}}
\end{figure*}

SiO masers of this type are commonly associated with evolved stars, or in very rare cases, star-forming molecular cores. As can be seen from Figure\,\ref{fig:SiO_Detection}, no counterpart is detected in CO(2-1) or CS(1-0) bands towards the SiO maser discovered in this investigation, discrediting the molecular core scenario.

%At infrared wavelengths (24, 8 and 5.8\,$\mu$m, Figure\,\ref{fig:RXJ_infrared}), a source coincident the SiO maser location is present.
Towards the SiO maser, one Midcourse Space Experiment (MSX) source (G347.0519$-$00.0100), 20 2MASS catalogue sources and 31 Spitzer-Glimpse catalogue sources are present within a 30$^{\prime\prime}$ radius\footnote{http://irsa.ipac.caltech.edu/}, so this maser may have an infrared counterpart (24, 8 and 5.8\,$\mu$m, Figures \ref{fig:RXJ_infrared} and \ref{fig:SiO_zoom}), but a more thorough investigation of this source is beyond the scope of this paper.
%The SiO maser may be associated with the coincident MSX source, G347.0519$-$00.0100, or possibly one of the 31 other Glimpse catalogue sources within 30$^{\prime\prime}$ of the maser, but an investigation of this source is beyond the scope of this research.

% but as yet, we have been unable to locate the source in Spitzer catalogues.

\section{Summary and Conclusion}
We carried out a 7\,mm survey of the northern and western regions of \RXJ , targeting the dense gas-tracing CS(1-0) transition and the shock-tracing SiO(1-0) transition. We report:

1) the discovery of dense components corresponding to Cores G and L at velocity, \vlsr$\sim -$11\,km\,s$^{-1}$, as well as gas in the Norma arm at \vlsr$\sim -$70\,km\,s$^{-1}$ and the 3\,kpc-expanding arm at \vlsr$\sim -$120\,km\,s$^{-1}$,

2) the discovery of a transient SiO(1-0) maser, in v=1 and 2 states, outside the western shell of the SNR, possibly generated by an evolved star,

3) the detection of dense gas and hot gas as indicated by CS(1-0) and CH$_3$OH(7-6)-emission coincident with the the infrared bubble S9 of \citet{Churchwell:2006},

4) and that a hadronic scenario for the RX\,J1713.7 $-$3946 SNR would require a CR enhancement above 1\,TeV of $\sim$1\,000, with dense CS(1-0)-traced components only contributing a gamma-ray flux above 1\,TeV of $\sim$3$\times$10$^{-13}$\,cm$^{-2}$\,s$^{-1}$, a few percent of the total hadronic flux. Systematic effects relating to the proportion of `hidden' gas and the assumed local CR spectrum may be $\sim$40\% and $\sim$40\%, respectively. %$\gtrsim$40\% isn't compiling!!!!

%In summary, we extend our Mopra 7\,mm survey of \RXJ\ and confirm the presence of dense gas towards Cores G and L (and background gas) via low levels of CS(1-0) emission. Complementary studies of SiO and CH$_3$OH emission may be of interest in investigating the activity associated with star formation.

%The \RXJ supernova remnant (SNR) is one of the brightest objects in the HESS TeV $\gamma$-ray sky and a good candidate studying the origin of galactic cosmic rays (CRs), which plausibly involves a SNR shock-gas interaction. 

\section*{Acknowledgments} %If needed
This work was supported by an Australian Research Council grant (DP0662810, DP1096533). The Mopra Telescope is part of the Australia Telescope and is funded by the Commonwealth of Australia for operation as a National Facility managed by the CSIRO. The University of New South Wales Mopra Spectrometer Digital Filter Bank used for these Mopra observations was provided with support from the Australian Research Council, together with the University of New South Wales, University of Sydney, Monash University and the CSIRO.

\begin{table*}[!h]
\centering
\small
\caption{The window set-up for the Mopra Spectrometer, MOPS, at 7\,mm. The centre frequency, targeted molecular line, targeted frequency and total mapping noise (T$_{\textrm{\tiny{RMS}}}$) are displayed.\label{Table:bands}} 
\begin{tabular}{|c|c|c|ccl|}
	\hline
Centre		& Molecular 					& Line		& \multicolumn{3}{c}{Map \Trms }\\ 
Frequency	& Emission Line					& Frequency & \multicolumn{3}{c}{(K\,ch$^{-1}$)}	\\
(GHz)		& 								& (GHz)		& West			& North & Northwest \\
	\hline
42.310 		& $^{30}$SiO(J=1-0,v=0) 		& 42.373365	& $\sim$0.07	&$\sim$0.07 &$\sim$0.10\\
42.500 		& SiO(J=1-0,v=3) 				& 42.519373	& $\sim$0.07	&$\sim$0.07 &$\sim$0.10\\
42.840 		& SiO(J=1-0,v=2) 				& 42.820582	& $\sim$0.07	&$\sim$0.08 &$\sim$0.10\\
			& $^{29}$SiO(J=1-0,v=0) 		& 42.879922	& 	& &\\
43.125 		& SiO(J=1-0,v=1) 				& 43.122079	& $\sim$0.08	&$\sim$0.07 &$\sim$0.12	\\
43.395 		& SiO(J=1-0,v=0) 				& 43.423864	& $\sim$0.08	&$\sim$0.09 &$\sim$0.12\\
44.085 		& CH$_3$OH(7(0)-6(1)\,A++) 	& 44.069476	& $\sim$0.08		&$\sim$0.10 &$\sim$0.12\\
45.125 		& HC$_7$N(J=40-39) 				& 45.119064	& $\sim$0.09	&$\sim$0.12 &$\sim$0.12	\\
45.255 		& HC$_5$N(J=17-16) 				& 45.26475	& $\sim$0.09	&$\sim$0.13 &$\sim$0.13\\
45.465 		& HC$_3$N(J=5-4,F=5-4) 			& 45.488839	& $\sim$0.09	&$\sim$0.13 &$\sim$0.13	\\
46.225 		& $^{13}$CS(J=1-0) 				& 46.24758	& $\sim$0.09	&$\sim$0.13 &$\sim$0.13\\
47.945 		& HC$_5$N(J=16-15) 				& 47.927275	& $\sim$0.12	&$\sim$0.15 &$\sim$0.15\\
48.225 		& C$^{34}$S(J=1-0) 				& 48.206946	& $\sim$0.12	&$\sim$0.15 &$\sim$0.15\\
48.635 		& OCS(J=4-3) 					& 48.651604	& $\sim$0.13	&$\sim$0.15 &$\sim$0.15\\
48.975 		& CS(J=1-0) 					& 48.990957	& $\sim$0.12	&$\sim$0.16 &$\sim$0.17\\
   \hline
\end{tabular}
\end{table*}

\begin{landscape}
%\begin{center}
\begin{table}
\caption{Parameters of CS(1-0) emission towards \RXJ. Galactic Coordinates, $[l$,$b]$, line-of-sight velocity, \vlsr, assumed distance, line-width (full-width-half-maximum), \fwhm, optical depth, average H column density, \Nh\,, and mass of H, $M$, are indicated.} Statistical uncertainties are on the order of 20\%, so errors are likely dominated by the systematics introduced by the analysis assumptions.\label{tab:Gas_Params}
\begin{tabular}{|l|c|c|c|c|c|c|c|}
\hline
Name$^a$ 	& Direction	&\vlsr 			& Assumed			&\fwhm 				& Optical 	& \Nh$^c$ 							& $M$ \\
			& $[l$,$b]$	& (km\,s$^{-1}$)& Distance (kpc)	& (km\,s$^{-1}$) 	& Depth$^b$ & ($\times$10$^{21}$\,cm$^{-2}$) 	& (M$_{\odot}$)	\\
%  & \tablehead{1}{r}{b}{Single\\outlet}
%  & \tablehead{1}{r}{b}{Small\tablenote{2-9 retail outlets}\\multiple}
%  & \tablehead{1}{r}{b}{Large\\multiple}
%  & \tablehead{1}{r}{b}{Total}   \\
%	&		&		& \\
\hline
Core\,A$^d$	&346.94$^{\circ}$,$-$0.32$^{\circ}$ & $-$9.82\,$\pm$\,0.02	&	1.0	& 1.25\,$\pm$\,0.06	& 0\,-\,0.44		& 3\,-\,4			& 12\,-\,15		\\
Core\,C$^d$ &347.08$^{\circ}$,$-$0.40$^{\circ}$ & $-$11.76\,$\pm$\,0.01	&	1.0	& 2.08\,$\pm$\,0.03	& 2.71\,$\pm$\,0.29	& 55\,$\pm$\,4	& 40\,$\pm$\,3 	\\
Core\,D$^d$ &347.31$^{\circ}$,0.01$^{\circ}$ &  $-$9.1\,$\pm$\,0.1	&	1.0	& 2.5\,$\pm$\,0.3	& 0\,-\,4.1			& 40\,-\,170		& 30\,-\,120	\\
Core\,G &347.033$^{\circ}$,$-$0.067$^{\circ}$ & $-$11.2\,$\pm$\,1.1 &	1.0	& 0.9\,$\pm$\,0.3		& 0\,-\,33$^e$			& 0.7\,-\,400$^e$		& 1\,-\,200$^f$\\	
Core\,L &347.433$^{\circ}$,$-$0.133$^{\circ}$ & $-$10.8\,$\pm$\,0.4	&	1.0	& 1.6\,$\pm$\,1.4		& 0\,-\,20$^e$			& 0.1\,-\,600$^e$			& 0.3\,-\,300$^f$\\
Clump\,N1-west&347.18$^{\circ}$,$+$0.01$^{\circ}$ & $-$70.1\,$\pm$\,0.2&6.0	& 4.6\,$\pm$\,0.5	& 0\,-\,4.7			& 130\,-\,610		& (2\,-\,10)$\times$10$^3$\\
Clump\,N1-east&347.24$^{\circ}$,$-$0.03$^{\circ}$ & $-$69.4\,$\pm$\,0.4&6.0	& 8.0\,$\pm$\,1.3	& 0\,-\,7.5			& 150\,-\,1\,100	& (3\,-\,20)$\times$10$^3$\\
Clump\,N3 &347.24$^{\circ}$,$+$0.02$^{\circ}$ & $-$76.6\,$\pm$\,0.3&6.0	 & 5.5\,$\pm$\,0.9	& 0\,-\,11	& 73\,-\,800	& (1\,-\,14)$\times$10$^3$\\
~~~N1+N3 region	&-	&-	&-	& - & -&	-& (5\,-\,30)$\times$10$^3$\\
Clump\,N2 &347.00$^{\circ}$,$-$0.01$^{\circ}$ & 	$-$70.9\,$\pm$\,0.2& 6.0		& 2.4\,$\pm$\,0.5	& 0\,-\,8.8			& 39\,-\,340		& (7\,-\,60)$\times$10$^2$\\
Clump\,T1-east &347.21$^{\circ}$,$-$0.09$^{\circ}$ & $-$120.0\,$\pm$\,0.2& 6.5		& 2.2\,$\pm$\,0.3	& 0\,-\,4.4			& 64\,-\,280		& (1\,-\,6)$\times$10$^3$\\
Clump\,T1-west &347.14$^{\circ}$,$-$0.07$^{\circ}$ & $-$119.5\,$\pm$\,0.1& 6.5	& 1.6\,$\pm$\,0.3	& 0\,-\,4.6			& 45\,-\,210		& (9\,-\,40)$\times$10$^2$\\
~~~T1 region	& -	&-&- & -&-	& -& (3\,-\,12)$\times$10$^3$\\
\hline
\end{tabular}
%\medskip
\\ \textit{$^a$Name convention from \citet{Moriguchi:2005} and this work.}
\textit{$^b$Assuming [CS]/[C$^{34}$S]$=$22.5 and using the C$^{34}$S(1-0)-band \Trms\ to estimate upper limits where no C$^{34}$S(1-0) emission is detected.}
\textit{$^c$Thermal Equilibrium (LTE) assumption at a temperature of 10\,K. Averaged over beam size.}
\textit{$^d$\citet{Maxted:2012}}
\textit{$^e$Lower values favoured (see \S\ref{sec:ResDis})}
\end{table}
%\end{center}
\end{landscape}


\begin{table*}[h]
\begin{center}
\caption{Line parameters for SiO transitions towards the detected SiO maser at [$l$,$b$]$\sim$[347.05,$-$0.012] for three observation periods (location indicated on Figure\,\ref{fig:CS_-10}). Background noise, \Trms, is shown. Velocity of peak, \vlsr , peak intensity, \Tpeak , and FWHM, \fwhm , were found by fitting Gaussians before deconvolving with the MOPS velocity resolution.
 \label{tab:SiO_Params}}
\begin{tabular}{|l|l|l|l|l|l|l|l|}
\hline
Date 		&SiO(1-0)	&\Trms 			&\vlsr 				& \Tpeak				& \fwhm 			\\
	 		&transition	&(K\,ch$^{-1}$)	&(km\,s$^{-1}$)		&						&(km\,s$^{-1}$) 	\\
%  & \tablehead{1}{r}{b}{Single\\outlet}
%  & \tablehead{1}{r}{b}{Small\tablenote{2-9 retail outlets}\\multiple}
%  & \tablehead{1}{r}{b}{Large\\multiple}
%  & \tablehead{1}{r}{b}{Total}   \\
%	&		&		& \\
\hline
			&v=0		&	0.09		&$^b-$4.9\,$\pm$\,0.3	&$^b$0.14\,$\pm$\,0.05	&$^b$1.5\,$\pm$\,0.7	\\	
23$/$24		&v=1		&	0.07		&$-$6.5\,$\pm$\,0.4		&0.11\,$\pm$\,0.03		&2.7\,$\pm$\,1.0	\\
April 2011	&v=2		&	0.08		&$-$6.3\,$\pm$\,0.1		&0.33\,$\pm$\,0.05		&2.6\,$\pm$\,0.7	\\
			&v=3		&	0.07		&	-					&	-					&	-				\\
\hline
			&v=0		&	0.11		&	-					&	-					&	-					\\
8			&v=1		&	0.08		&$^a-$4.2\,$\pm$\,0.3	&$^a$0.08\,$\pm$\,0.02	&$^a$3.9\,$\pm$\,2.5	\\
April 2012	&v=2		&	0.07		&$^a-$4.7\,$\pm$\,0.6	&$^a$0.10\,$\pm$\,0.02	&$^a$5.2\,$\pm$\,0.8	\\
			&v=3		&	0.06		&	-					&	-					&	-				\\
\hline		
			&v=0		&	0.03		&	-					&	-					&	-				\\
20			&v=1		&	0.03		&$-$4.8\,$\pm$\,0.3		&0.07\,$\pm$\,0.01		&3.9\,$\pm$\,0.6	\\
April 2012	&v=2		&	0.03		&$-$5.0\,$\pm$\,0.3		&0.083\,$\pm$\,0.009	&5.2\,$\pm$\,0.6	\\
			&v=3		&	0.03		&	-					&	-					&	-				\\
			\hline	
\end{tabular}
\\\textit{$^a$Spectral lines difficult to fit, so line widths were fixed to be equal to those of the the April 20, 2012 observations. 
$^b$Line deemed insignificant in post-trial analysis (see \S\ref{sec:SiO_2011})} 
%\medskip
\end{center}
\end{table*}

\appendix

\begin{figure*}%[!h]
\centering
\includegraphics[width=0.99\textwidth]{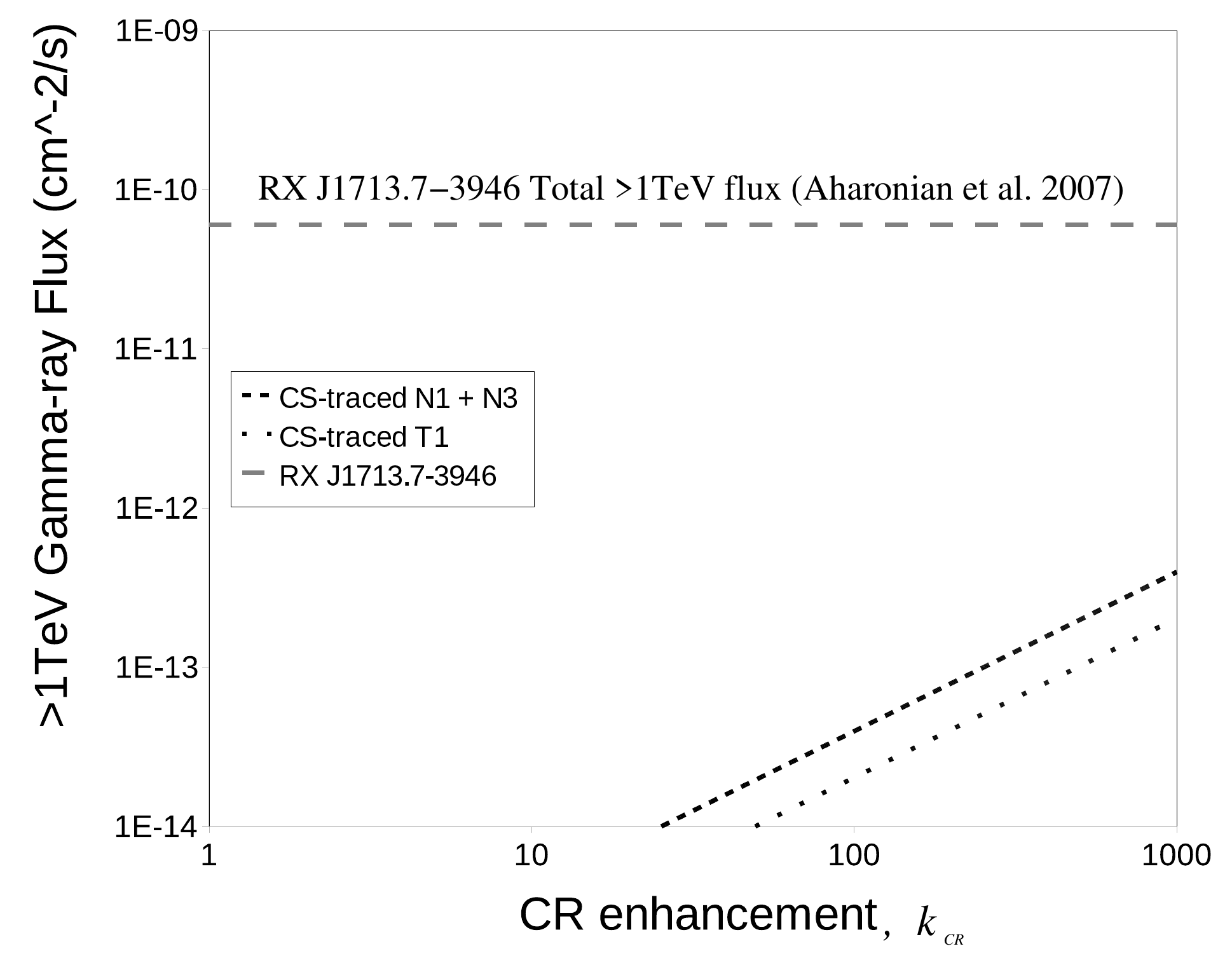}
\caption{The predicted hadronic $\gamma$-ray flux as a function of CR-enhancement if dense gas currently believed to be background to \RXJ\ (\vlsr$\sim -$70, $-$120\,km\,s$^{-1}$), were in fact, associated with \RXJ . Note that we only include contributions from CS(1-0) emission. Distances of 6\,kpc and 6.5\,kpc were assumed for N1/3 and T1, respectively.\label{fig:CRenhanceBG}}
\end{figure*}

%\begin{figure}%[!h]
%\centering
%\includegraphics[width=0.49\textwidth]{Paper3_pics/AlternateImages}
%\caption{Alternate CS(1-0) images that include HESS contours. \label{fig:AltIm}}
%\end{figure}

\begin{figure*}
\centering
\includegraphics[width=0.49\textwidth]{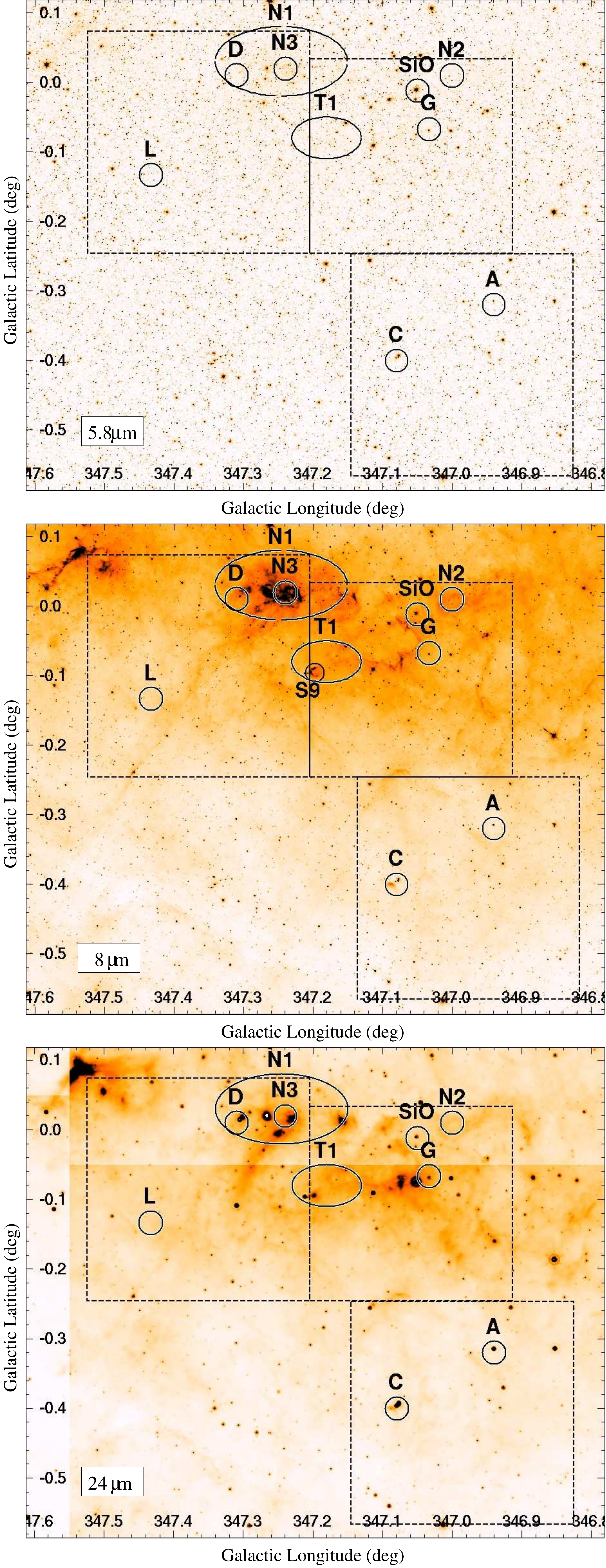}
\caption{Spitzer 5.8\,$\mu$m (top), 8\,$\mu$m (middle) and 24\,$\mu$m (bottom) emission images. Objects of interest to this investigation (see \S\ref{sec:CS}) are indicated by circles and ellipses. Object S9 towards T1 (see text) is indicated in the middle picture (8\,$\mu$m). Dashed squares indicate regions mapped in 7\,mm wavelengths by Mopra.\label{fig:RXJ_infrared}}
\end{figure*}

\begin{figure*}
\centering
\includegraphics[width=0.49\textwidth]{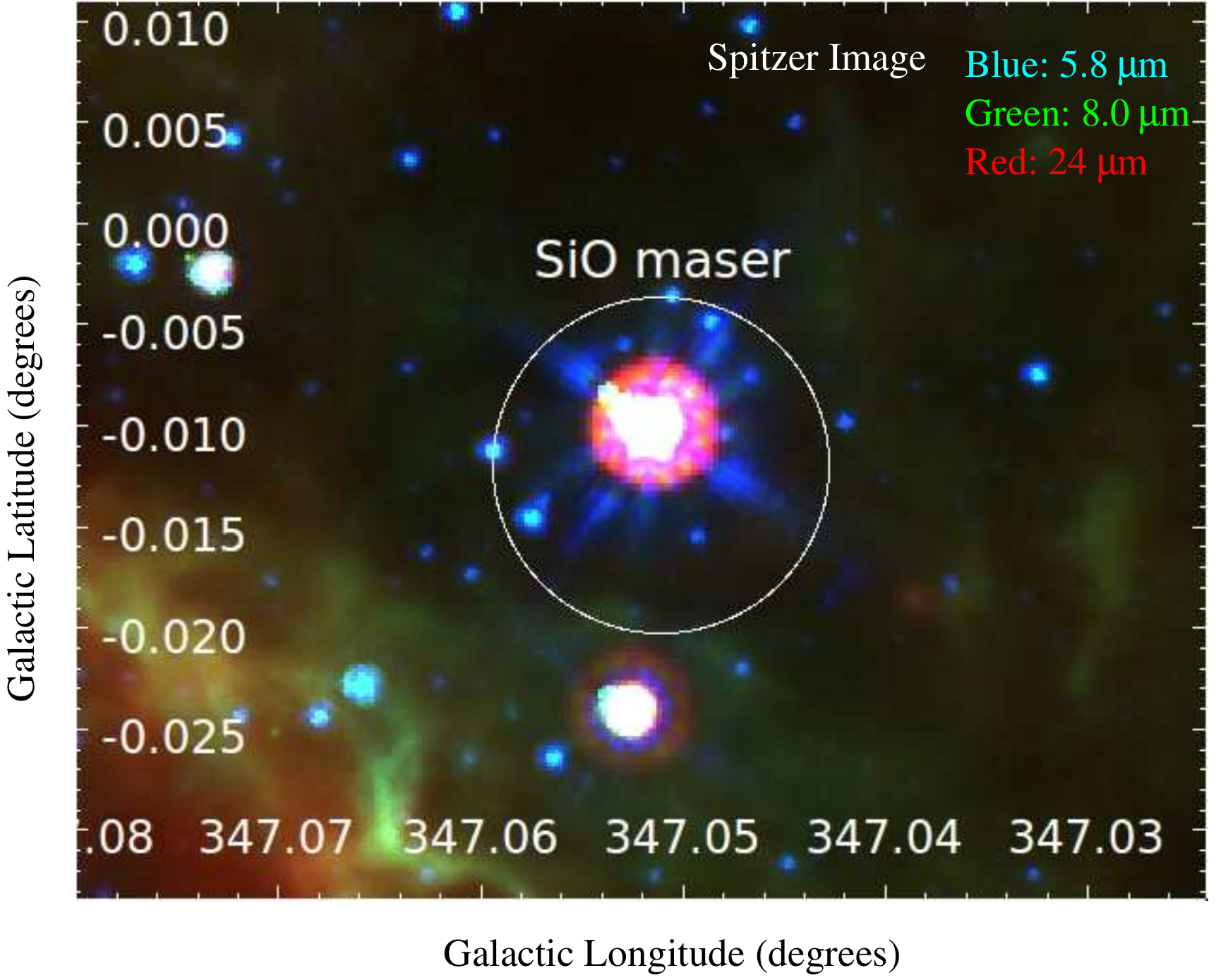}
\caption{Three colour image of Spitzer 5.8\,$\mu$m, 8\,$\mu$m and 24\,$\mu$m emission towards the SiO maser discovered in our survey. The Mopra 7\,mm beam FWHM is displayed (white circle). \label{fig:SiO_zoom}}
\end{figure*}

%%Format tables as in the following example
%\begin{table}[h]
%\begin{center}
%\caption{Example Table}\label{tableexample}
%\begin{tabular}{lcc}
%\hline Column 1 & Column 2 & Column 3 \\
%\hline Table Content$^a$ \\
%\hline
%\end{tabular}
%\medskip\\
%$^a$Table footnotes go here.\\
%\end{center}
%\end{table}

\end{document}